\documentclass[12pt, a4paper]{article}

\usepackage[utf8]{inputenc}
\usepackage{amssymb,amsmath,amsfonts}
\usepackage{eurosym,geometry,ulem}
\usepackage{graphicx}
\usepackage{color,setspace,sectsty,comment,footmisc}
\usepackage[authoryear]{natbib}
\usepackage{subfig} 
\usepackage{tabularx,booktabs,array,multirow}
\usepackage{float}
\usepackage{caption}  
\usepackage{url}
\usepackage{hyperref}  
\usepackage{adjustbox}
\usepackage{lscape}
\usepackage{epsfig}
\usepackage{dcolumn}
\usepackage{longtable,dcolumn,tabularx}
\usepackage{exscale,amsthm,multirow,rotating}
\usepackage{bbm,soul}
\usepackage{tikz,dsfont,float}
\usepackage{subcaption}
\usepackage{afterpage}
\usepackage{endnotes}
\let\footnote=\endnote 

\hypersetup{
    colorlinks=true,
    linkcolor=blue,
    citecolor=blue,
    filecolor=magenta,      
    urlcolor=cyan,
    pdftitle={review paper},
    pdfpagemode=FullScreen,
}

\newcolumntype{L}[1]{>{\raggedright\let\newline\\arraybackslash\hspace{0pt}}m{#1}}
\newcolumntype{C}[1]{>{\centering\let\newline\\arraybackslash\hspace{0pt}}m{#1}}
\newcolumntype{R}[1]{>{\raggedleft\let\newline\\arraybackslash\hspace{0pt}}m{#1}}

\begin{document}

\begin{titlepage}

\title{Forecasting Oil Volatility through Network Models with GARCH-Informed Correlation Weights}
\author{Fayçal Djebari\thanks{Department of Economics, University of Bejaia, Algeria. Email: \href{mailto:faycal.djebari@univ-bejaia.dz}{faycal.djebari@univ-bejaia.dz}; \\ ORCID: \href{https://orcid.org/0009-0002-9265-9541}{0009-0002-9265-9541}}\and 
	Kahina Mehidi\thanks{Department of Economics, University of Bejaia, Algeria. Email: \href{mailto:kahina.mehidi@univ-bejaia.dz}{kahina.mehidi@univ-bejaia.dz}; ORCID: \href{https://orcid.org/0009-0001-3953-5155}{0009-0001-3953-5155}} \and 
	Khelifa Mazouz\thanks{Department of Economics, University of Bejaia, Algeria. Email: \href{mailto:khelifa.mazouz@univ-bejaia.dz}{khelifa.mazouz@univ-bejaia.dz}; \\ Cardiff Business School, Cardiff University, United Kingdom. Email: \href{mailto:MazouzK@cardiff.ac.uk}{MazouzK@cardiff.ac.uk}; ORCID: \href{https://orcid.org/0000-0001-6711-1715}{0000-0001-6711-1715}} \and 
	Philipp Otto\thanks{School of Mathematics and Statistics, University of Glasgow, United Kingdom. \\ Email: \href{mailto:Philipp.Otto@glasgow.ac.uk}{Philipp.Otto@glasgow.ac.uk}; ORCID: \href{https://orcid.org/0000-0002-9796-6682}{0000-0002-9796-6682}}}
\date{\today}
\maketitle
\vspace{-0.3in}

\begin{abstract}

\noindent This study addresses the computational challenges of forecasting volatility in high-dimensional commodity markets. Building on the Network log-ARCH framework, we introduce a novel class of network topologies from GARCH-informed correlation weights, obtained from conditional covariance estimates of multivariate GARCH models, rather than relying on the heuristic distance measures commonly used in clustering methods. We evaluate the proposed models forecasting performance through a rolling-window exercise using a panel of OPEC members crude oil prices. The results identify network volatility models incorporating these new GARCH-informed weights as the statistically superior specifications. Remarkably, the proposed framework matches standard DCC-GARCH predictive accuracy while delivering up to 62,000-fold computational gains. By explicitly modeling contemporaneous spillovers through interpretable spatial ARCH-like lags estimated via GMM, the proposed approach offers an optimal trade-off between parsimony, interpretability, and performance. The findings establish GARCH-informed network models as robust, scalable alternatives for systemic risk measurement and volatility forecasting in interconnected financial markets.\\
\vspace{-0.1in}
\vspace{-0.1in}

\end{abstract}
\setcounter{page}{0}
\thispagestyle{empty}
\end{titlepage}
\pagebreak \newpage

\onehalfspacing

 \section{Introduction}
  \label{sec1}

   Volatility forecasting poses a fundamental challenge across commodity markets, with petroleum markets showing exceptional complexity. The crude oil market ranks among the most volatile commodity markets \citep{regnierOilEnergyPrice2007}, and one of the most challenging assets to forecast. This volatility calls for innovative forecasting strategies and a deep understanding of price fluctuation mechanisms, which are crucial for practitioners and policymakers managing significant economic risks. The Organization of the Petroleum Exporting Countries (OPEC), one of the most influential organizations in global oil markets, plays a central role in market stabilization by coordinating production and pricing policies among member nations. Acting as a cartel, OPEC restricts oil supply to support price levels \citep{kilianNotAllOil2009, AlRousan2018}. Despite this coordination, significant heterogeneity persists across member countries regarding production behavior and pricing dynamics, implying complex interdependencies within the cartel structure. As a result, volatility transmission within OPEC may be highly non-trivial. Understanding these interdependencies is therefore critical for accurate volatility forecasting in global oil markets.

   One promising approach to improving forecasts in interconnected markets is through network frameworks. Financial markets can be effectively represented as networks where asset returns serve as nodes and edges reflect the degree of similarity or connection between them \citep[e.g.][]{Barigozzi2016,Betancourt2019,Demiris2013,Diebold2014,Vinciotti2019,Zhou2023}. Recent studies have increasingly adopted these network-based methodologies to explore volatility spillovers and market interconnectedness \citep[e.g.][]{AlRousan2018, Ji2018, Barunk2018, Barigozzi2019, Bykhovskaya2021, Dai2023, Uddin2023, Barigozzi2025}. Many of these models build upon the connectedness framework of \citet{Diebold2009,Diebold2012,Diebold2014}, which quantifies directional spillovers using forecast error variance decompositions. However, these approaches often impose structure through rolling-window VARs and do not directly model the volatility process or incorporate the endogenous network structure of conditional heteroskedasticity, thereby limiting their ability to capture the structural mechanisms governing volatility transmission in highly interconnected markets.
   
   \citet{matteraNetworkLogARCHModels2024} address this limitation by introducing the Network Log-ARCH framework, which embeds network structure directly into the volatility process through spatial econometric methods. Their model captures contemporaneous volatility spillovers, where shocks in one market instantly affect volatility in connected markets, while maintaining computational tractability through GMM estimation. Applying this framework to stock market forecasting, they demonstrate superior performance over independent univariate log-ARCH models, with recent evidence suggesting that network-informed models can lead to better forecasts in financial markets \citep[e.g.][]{Wu2022,Huang2023}. However, their analysis relies primarily on unconditional proximity measures, such as Euclidean distance and simple Pearson correlation, alongside K-nearest neighbors, to construct network weights. This approach does not leverage the rich information embedded in the conditional variance-covariance structure of financial returns, leaving open the question of whether economically informed network topologies could further enhance forecasting accuracy. Moreover, while several studies analyze volatility connectedness across oil markets, few have embedded empirical network structures directly into the volatility generating process \citep[e.g.][]{Bauwens2006,Ashfaq2020,Wang2021,SnchezGarca2023,BenSalem2024}.

  This study addresses these gaps by extending the Network Log-ARCH framework in two key directions. First, we construct network weight matrices derived from the conditional correlation structures of standard MGARCH models, specifically CCC-GARCH, DCC-GARCH, and GO-GARCH. Unlike unconditional distance measures, these GARCH-informed networks encode the correlation structure implied by the underlying volatility dynamics of the system, capturing economic dependencies that purely geometric or unconditional statistical measures fail to detect. Importantly, we use time-averaged weights computed dynamically from the in-sample conditional correlations, providing stable network structures that reflect structural interdependencies without introducing look-ahead bias. Second, we apply this enhanced framework to OPEC oil markets, where production coordination and policy heterogeneity create unique spillover patterns distinct from equity markets. Building on the spatiotemporal ARCH modeling literature \citep{ottoDynamicSpatiotemporalARCH2024}, our approach enables modeling of structural volatility interdependence across a network of interconnected economies while directly inferring spillover dynamics from historical log return data.
  
  Accordingly, this paper contributes to the literature in three main ways. First, it introduces GARCH-informed network weight matrices into the Network log-ARCH specification, derived from CCC, DCC, and GO-GARCH conditional correlations, allowing the model to exploit economically meaningful dependence structures. Second, it applies this framework to a panel of OPEC oil-exporting economies, providing new insights into the transmission of volatility among these interconnected member countries and demonstrating that network topology is a key determinant of forecasting performance. Third, it evaluates forecasting accuracy against standard multivariate GARCH specifications, demonstrating that GARCH-informed network models achieve comparable or superior predictive performance while drastically reducing computational cost, thereby offering a scalable alternative to traditional high-dimensional volatility models.
  
  Our empirical findings confirm these advantages. Network Log-ARCH models incorporating CCC-, DCC-, and GO-GARCH-derived weight matrices yield lower Root Mean Squared Forecast Errors (RMSFE) and Mean Absolute Forecast Errors (MAFE) than Euclidean and correlation-based network alternatives, and perform competitively with, or better than, standard multivariate GARCH specifications. Predictive accuracy is evaluated using the \citet{Diebold2002} test and its heteroscedasticity-robust extension \citep{Harvey2024}. Furthermore, the best-performing models are rigorously identified through the Model Confidence Set (MCS) procedure of \citet{hansenModelConfidenceSet2011}. Notably, the three GARCH-informed network specifications are jointly retained in the MCS at the 5\% level and produce nearly identical point forecasts, while simpler network topologies (Euclidean, Pearson correlation, Piccolo distance) are excluded. These results highlight the central role of conditional correlation structures in shaping volatility dynamics across interconnected markets.

  In the context of global energy markets, these models equip academic and professional energy economists, as well as policymakers, with a precise mechanism for tracking how price shocks propagate across major oil-exporting nations through contemporaneous spillovers. The network weights, particularly those derived from GARCH-informed conditional correlations, capture structural patterns in market interconnectedness. Crucially, this computational advantage transforms market surveillance. By shifting from retrospective to near real-time monitoring, it gives authorities a vital window to anticipate volatility shocks before they spread across the macroeconomic network. For economies heavily dependent on hydrocarbon revenues, such as OPEC member states, this enhanced volatility forecasting capability has direct implications for sovereign wealth management, macroeconomic planning, and the mitigation of risks stemming from systemic energy market shocks. Furthermore, these models provide practitioners and investors with highly efficient, advanced tools that drastically accelerate the forecasting process in network settings, supporting timely and informed financial decision-making under uncertainty.

  The remainder of this paper is organized as follows. Section~\ref{sec2} reviews the relevant literature on oil price volatility, forecasting, and network-based modeling. Section~\ref{sec3} describes the dataset and presents the key descriptive statistics. Section~\ref{sec4} provides an overview of multivariate volatility models. Section~\ref{sec5} outlines the methodological framework for the network-based log-ARCH models and the construction of the network weight matrices. Section~\ref{sec6} presents the empirical findings, including the estimation results, network structures highlighting interconnectedness and spillover effects, and the comparative forecasting performance of the proposed models; it concludes by outlining the advantages of using the proposed network volatility models.

  \section{Literature Review}
  \label{sec2}

  Since its establishment in Baghdad in 1960, OPEC has aimed to coordinate and harmonize the petroleum policies of its member countries to ensure stable oil prices and adequate returns on investment \citep{OPECstatute}. Over the decades, the organization has been shaped by a series of major historical and geopolitical events \citep{opecOrganizationPetroleum}, many of which have had far-reaching effects on the global economy \citep{EffetsOilPriceGlobalEco}. These events have exposed the persistent susceptibility of oil-exporting countries to external shocks, highlighting the complexity of managing coordinated policies in the face of heterogeneous economic responses \citep{LIU2019104548}. Geopolitical risks, particularly in oil-producing regions such as the Middle East and North Africa, have consistently driven oil price volatility \citep{Mignon2024}. For instance, regional tensions and conflicts, along with global crises—such as the Chinese economic slowdown \citep{hamiltonChangingFaceWorld2014}—have led to sharp fluctuations in oil prices. More recently, the COVID-19 pandemic amplified this volatility by triggering both supply- and demand-side shocks \citep{bourghelleOilPriceVolatility2021, LE2023106474}. Such compounded disruptions underscore the importance of understanding how volatility transmits across oil-exporting economies, particularly within policy-coordinated networks such as OPEC, where shocks may propagate unevenly across members.

  Economies that are heavily dependent on oil exports face considerable challenges due to persistent shocks and uncertainties in global oil markets. These fluctuations hinder long-term economic planning, disrupt investment strategies, and constrain economic growth \citep{Sachs1995}. To mitigate these adverse effects, it is crucial for such countries to diversify their energy sources and reduce their reliance on oil revenues, thereby limiting their exposure to oil price volatility \citep{kilianEconomicEffectsEnergy2008, kilianOilPriceShocks2014}. Diversification enhances economic resilience by insulating national economies from the destabilizing effects of external shocks.

  Oil price volatility remains a central concern in economic and financial forecasting, given its influence on key macroeconomic indicators, such as investment, durable goods consumption, and aggregate output. Uncertainty generated by oil price fluctuations can lead to reduced economic activity, particularly in resource-rich exporting nations \citep{Finn2000Perfect, kilianEconomicEffectsEnergy2008, elderOilPriceUncertainty2010, VO2021105391}. These countries, as emphasized by \citet{CAPPELLI2023113479}, are disproportionately affected due to their strong fiscal reliance on oil revenues. Despite decades of research, accurately forecasting oil price movements remains a formidable task due to complex interactions among market agents \citep{baumeisterFortyYearsOil2016}.

  Volatility in oil markets also has profound implications for corporate investment behavior. Elevated uncertainty about future energy input costs, as discussed by \citet{pindyckIrreversibilityUncertaintyInvestment1991}, deters firms from undertaking irreversible investment projects. This effect is more pronounced in oil-exporting economies, where future profitability is directly tied to oil price movements. Numerous empirical studies corroborate the negative relationship between oil price uncertainty and investment levels \citep{henriquesEffectOilPrice2011, rattiRelativeEnergyPrice2011, wangInternationalOilPrice2017}. The exceptional volatility of oil and gas prices, surpassing that of most commodities, is attributed to supply-demand dynamics, macroeconomic fluctuations, and geopolitical tensions, as evidenced by research papers such as \citep{regnierOilEnergyPrice2007} and \citep{demirbasRecentVolatilityPrice2017}.

  Conventional econometric models often fail to fully capture the complexity of oil price volatility, particularly its dynamic and interconnected nature within coordinated economic blocs. In response, \citet{engleAutoregressiveConditionalHeteroscedasticity1982} and \citet{bollerslevGeneralizedAutoregressiveConditional1986} developed the ARCH and GARCH models, which model volatility clustering in financial time series. These nonlinear frameworks remain foundational in financial econometrics, especially through multivariate extensions such as CCC-GARCH \citep{Bollerslev1990}, DCC-GARCH \citep{Engle2002}, and GO-GARCH \citep{vanderWeide2002}, which accommodate dynamic correlations in the return series. More recently, attention has shifted to models that incorporate spatial and network-based dependencies to account for cross-country volatility transmission. Spatiotemporal ARCH (spARCH) models incorporate contemporaneous spatial and temporal dependence to better capture volatility interlinkages \citep{ottoGeneralizedSpatialSpatiotemporal2018}, while spatial and spatiotemporal log-ARCH models \citep{satoSpatialExtensionGeneralized2021, ottoDynamicSpatiotemporalARCH2024} and network log-ARCH models \citep{matteraNetworkLogARCHModels2024} extend this framework to account for multiplicative effects. A comprehensive synthesis of these advanced approaches is presented in \citet{ottoSpatialSpatiotemporalVolatility2024}.

  Empirical studies have shown strong interdependencies between oil prices and financial markets \citep{AWARTANI201328, GUHATHAKURTA2020104566, GUAN2024107305}. For instance, \citet{nguyenOilPriceDeclines2020} document the short-run global repercussions of oil price shocks, while \citet{cretiOilPriceFinancial2014} observe heightened oil–stock market co-movements in oil-exporting countries. However, existing studies often rely on models that assume static or exogenous relationships, limiting their ability to reflect the endogenous and evolving nature of the interdependencies among oil-exporting countries. In response, recent advances—such as time-series clustering methods \citep{Maharaj2019} embedded into Dynamic Spatiotemporal ARCH models \citep{ottoDynamicSpatiotemporalARCH2024}—offer promising tools for inferring network structures directly from return data. This opens new avenues for modeling volatility transmission in a way that reflects both temporal dynamics and cross-sectional spillovers, which is particularly relevant for OPEC economies, where coordinated production masks heterogeneous national responses. By integrating clustering-derived networks with advanced volatility models, our approach captures the latent and evolving interdependencies that shape volatility spillovers in oil-exporting nations. These methods move beyond traditional frameworks by eliminating the need for external variables, allowing for real-time analysis rooted in the structure of the return data itself. This offers both improved forecast accuracy and deeper insights into the underlying transmission channels relevant for policy coordination and risk management.

   \section{Data, Descriptive Statistics and Preliminary Analyses}
  \label{sec3}
  
  The empirical analysis in this study is based on monthly crude oil price data for six key OPEC member countries: Algeria (Saharan Blend), Iran (Heavy), Libya (Es Sider), Nigeria (Bonny Light), Saudi Arabia (Arab Light), and the United Arab Emirates (Murban). These countries were selected based on the availability of a complete and consistent historical time series, which offered the same number of monthly observations across the entire sample. Crude oil prices were retrieved from the official OPEC database and subsequently converted into monthly logarithmic returns. The dataset spans the period from January 1983 to December 2024.
  
\begin{figure}[h!]
    \centering
    \subfloat[
    \centering
    ]{{\includegraphics[width=7cm]{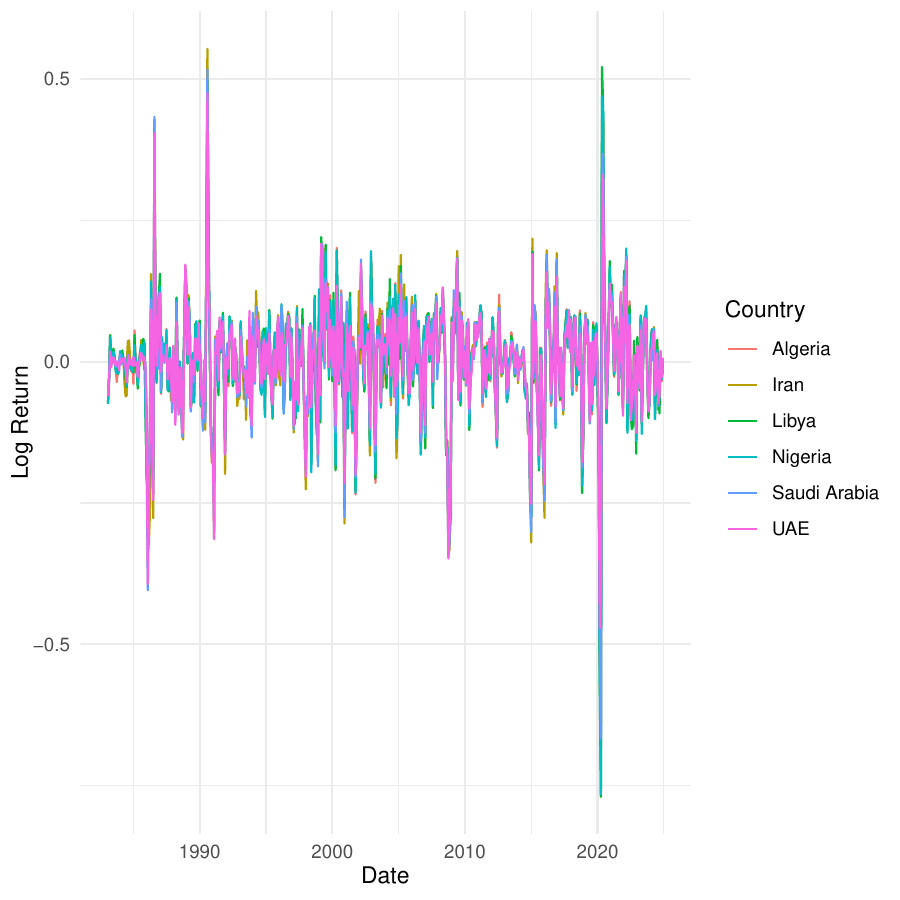} }}%
    \qquad \subfloat[
    \centering
    ]{{\includegraphics[width=7cm]{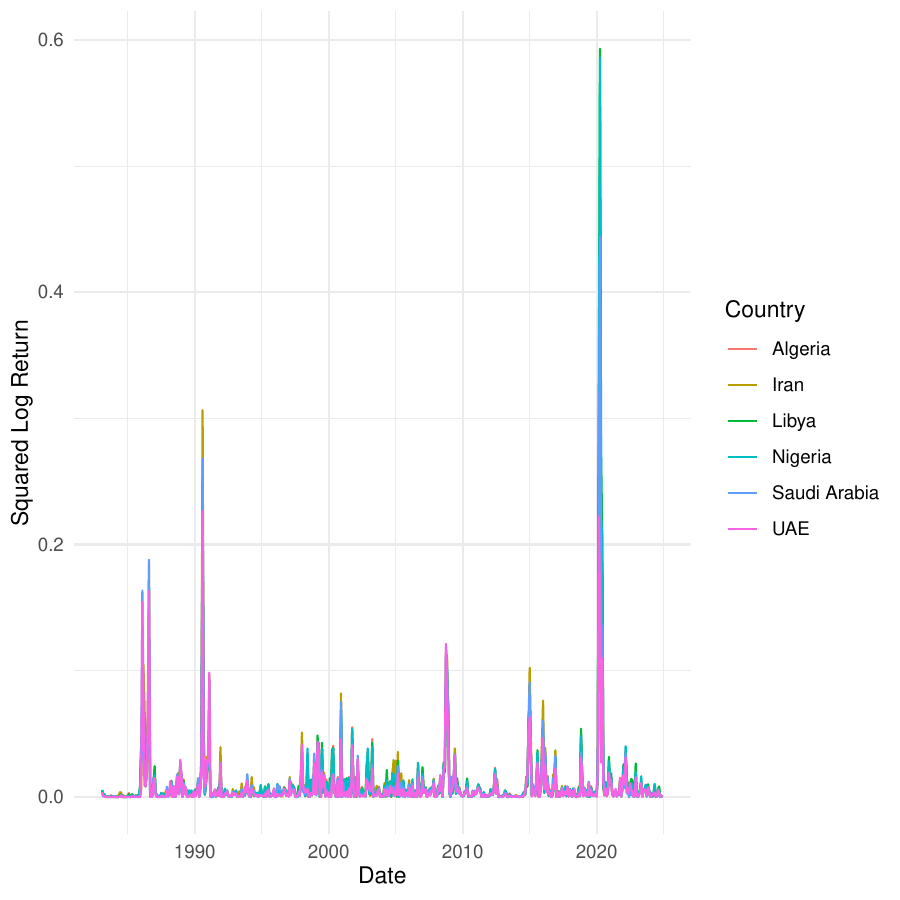} }}%
    \caption{Panel (a) displays the monthly log returns and panel (b) shows the squared monthly log returns of oil prices from January 1983 to December 2024 for six selected OPEC countries: Algeria, Iran, Libya, Nigeria, Saudi Arabia, and the United Arab Emirates.}%
    \label{fig:graph}
  \end{figure}

  The logarithmic returns and their squared values are presented in Figure~\ref{fig:graph}, which illustrates the volatility dynamics of the OPEC crude oil markets over time. Notable spikes in volatility coincide with major global events and financial disruptions, including the 1986 oil price collapse, the Gulf War in 1990, the 2008 global financial crisis, the 2014 oil price downturn \citep{Baumeister2016,Prest2018}, the COVID-19 pandemic in 2020 \citep{bourghelleOilPriceVolatility2021, LE2023106474}, and the Russia–Ukraine conflict \citep{ZHANG2023106956}. These episodes, which triggered pronounced fluctuations in oil prices, are consistent with the historical patterns of oil shocks documented by \citet{hamiltonHistoricalOilShocks2011}.

  The descriptive statistics for the monthly log returns of prices across the six selected OPEC producers are presented in Table~\ref{tab:stats}. All series exhibit positive means and medians, reflecting general upward trends but with varying dispersion levels, suggesting heterogeneous volatility. Negative skewness and excess kurtosis are present in all cases, indicating asymmetric distributions with fat tails and substantial tail risk. These features—non-normality, asymmetry, and leptokurtosis—justify the use of heavy-tailed distributions in subsequent modeling.
  
  The distributional plots in Appendix Figures~\ref{fig_app:density_plots} and~\ref{fig_app:qq_plots} complement these findings by illustrating the empirical deviations. While returns are broadly centered around zero, the dispersion and shape vary markedly across countries. Libya exhibits the widest dispersion and the most extreme heavy tails, indicating a higher susceptibility to extreme shocks. Nigeria also displays a broad spread and the most pronounced asymmetry among the group. In contrast, Saudi Arabia and the UAE show tighter and more concentrated return profiles, though they still exhibit significant deviations from the Gaussian benchmark. All distributions decisively reject normality, as confirmed by the \citet{Jarque1980} tests in Table~\ref{tab:stats}. These stylized facts are consistent with the well-documented empirical properties of financial returns \citep{Bera1993}.

\begin{table}[!htbp]
	\centering
	\caption{Descriptive Statistics of Monthly Oil Price Log Returns}
	\label{tab:stats}
	\small 
	\setlength{\tabcolsep}{3.5pt} 
	\renewcommand{\arraystretch}{1.2} 
	\begin{tabular}{lrrrrrrrr}
		\toprule
		\textbf{Country} & \textbf{Mean} & \textbf{Median} & \textbf{Std. Dev.} & \textbf{Min.} & \textbf{Max.} & \textbf{Skew.} & \textbf{Kurt.} & \textbf{JB $p$-value} \\
		\midrule
		Algeria      & 0.0017 & 0.0066 & 0.0999 & $-0.6952$ & 0.4957 & $-0.683$ & 8.04 & $<$2.2e-16*** \\
		Iran         & 0.0019 & 0.0053 & 0.1028 & $-0.6475$ & 0.5536 & $-0.605$ & 6.82 & $<$2.2e-16*** \\
		Libya        & 0.0017 & 0.0071 & 0.1039 & $-0.7700$ & 0.5215 & $-0.798$ & 9.89 & $<$2.2e-16*** \\
		Nigeria      & 0.0017 & 0.0088 & 0.1019 & $-0.7654$ & 0.4708 & $-0.838$ & 9.67 & $<$2.2e-16*** \\
		Saudi Arabia & 0.0018 & 0.0054 & 0.0975 & $-0.6665$ & 0.5176 & $-0.826$ & 8.36 & $<$2.2e-16*** \\
		UAE          & 0.0017 & 0.0074 & 0.0893 & $-0.4718$ & 0.4757 & $-0.612$ & 5.79 & $<$2.2e-16*** \\
		\bottomrule
	\end{tabular}
	\vspace{1ex}
	\begin{minipage}{0.95\textwidth}
		\footnotesize
		\textbf{Notes:} JB refers to the Jarque--Bera test for normality. *** indicates $p < 0.001$. All series show non-normality with negative skewness and excess kurtosis.
	\end{minipage}
\end{table}

    \subsection{Interdependence of Oil Price Returns Among OPEC Countries}
  
  The correlation matrix in Table~\ref{tab:correlation} shows strong positive comovement among the log returns of oil prices across the six OPEC countries, indicating high market integration and reflecting OPEC's role in market stabilization. The highest correlations appear among the African members, particularly between Algeria, Nigeria, and Libya, likely because of their similar exposure to global oil shocks. Gulf countries show a distinct pattern, as Saudi Arabia and the UAE correlate more strongly with each other and with Iran than with the African members, indicating a degree of intra-Gulf cohesion. In contrast, slightly weaker correlations for Iran and the UAE with the North African countries may stem from geopolitical factors or divergent production policies. These findings suggest that while global shocks affect all members similarly, regional asymmetries persist, supporting the use of both constant and dynamic correlation models, such as CCC and DCC-GARCH \citep{Bollerslev1990,Engle2002}, as well as network-based approaches inspired by spatiotemporal ARCH models \citep[see][]{ottoSpatialSpatiotemporalVolatility2024}.

\begin{table}[h!]
	\centering
	\caption{Correlation Matrix of Monthly Oil Price Log Returns}
	\label{tab:correlation}
	\begin{tabular}{lcccccc}
		\hline
		& Algeria & Iran   & Libya  & Nigeria & Saudi Arabia & UAE    \\
		\hline
		Algeria      & 1  &  &  &   &      &  \\
		Iran         & 0.9513  & 1 & &   &       &  \\
		Libya        & 0.9927  & 0.9513 & 1 &   &        & \\
		Nigeria      & 0.9945  & 0.9522 & 0.9932 & 1  &      &  \\
		Saudi Arabia & 0.9526  & 0.9806 & 0.9534 & 0.9566  & 1      &  \\
		UAE          & 0.9378  & 0.9677 & 0.9354 & 0.9360  & 0.9769       & 1 \\
		\hline
	\end{tabular}
\end{table}

\section{Overview of Time-Series Volatility Models}
  \label{sec4} 
  
  This section reviews the multivariate specifications—CCC, DCC, and GO-GARCH—that capture static and dynamic conditional correlations across assets, and are subsequently used to construct model-based weight matrices for the Network log-ARCH model.

  \subsection{Multivariate GARCH models}
  \label{subsec43} 
  
 Multivariate GARCH models are designed to capture the joint dynamics and cross-dependencies across assets. \citet{bollerslev1988capital} introduced the VEC-GARCH model, which generalizes the GARCH framework to the multivariate setting by allowing time-varying covariance and correlation structures. Let \( \mathbf{Y}_t \) be an \( r \)-dimensional random vector of asset returns. Then:
  \begin{equation}
    \label{eq:vecgarch}
     \mathbf{Y}_t = \mathbf{H}_t^{1/2} \varepsilon_t,
  \end{equation}
  where \( \varepsilon_t \) is an i.i.d. innovation vector and \( \mathbf{H}_t = \operatorname{Var}(\mathbf{Y}_t \mid \mathcal{F}_{t-1}) \) is the conditional covariance matrix, with:
  \begin{equation}
    \label{eq:VarHt}
    \text{Var}(\mathbf{Y}_t \mid \mathbf{Y}_{v}, v < t) = \mathbf{H}_t.
  \end{equation}

  Various multivariate GARCH models address the growing complexity of estimating \( \mathbf{H}_t \) as the number of series increases. In this study, we focus on the CCC-GARCH, DCC-GARCH, and GO-GARCH models, each of which offers distinct features and levels of flexibility. These specifications are particularly suited for constructing model-based weight matrices in our network analysis of monthly log returns from the six OPEC countries. The following subsections discuss these models in increasing order of structural flexibility, highlighting their methodological linkages.

  \subsubsection{Constant and Dynamic Conditional Correlation GARCH Models}
  \label{subsec421} 
  
  To capture co-movement in financial returns with time-varying volatility, \citet{Bollerslev1990} proposed the Constant Conditional Correlation (CCC) GARCH model. This framework assumes that, while individual asset volatilities may evolve over time, their correlations remain constant. Formally, the conditional covariance matrix \( \mathbf{H}_t \) is decomposed as:
  \begin{equation}
    \label{eq:ccc_Ht}
    \mathbf{H}_t = \mathbf{D}_t \mathbf{R} \mathbf{D}_t,
  \end{equation}
  where \( \mathbf{R} \) is a fixed correlation matrix, and \( \mathbf{D}_t \) is a diagonal matrix containing conditional standard deviations,
  \begin{equation}
    \label{eq:cond_std_dev}
    \mathbf{D}_t = \text{diag}(\sqrt{h_{1t}}, \ldots, \sqrt{h_{Nt}}).
  \end{equation}
  
  Each \( h_{it} \) follows a univariate GARCH$(p,q)$ process, as defined in \eqref{eq:univ_garch}. Its general form is:
  \begin{equation}
    h_{it} = \alpha_0^{(i)} + \sum_{j=1}^{p} \alpha_j^{(i)} Y_{i,t-j}^2 + \sum_{j=1}^{q} \beta_j^{(i)} h_{i,t-j}, \quad i = 1, \ldots, N.
  \end{equation}

  While CCC-GARCH offers a simple yet effective structure, the assumption of constant correlations can be overly restrictive. To address this, \citet{Engle2002} introduced the Dynamic Conditional Correlation (DCC) model, which preserves the same matrix form,
  \begin{equation}
    \label{eq:dcc_Ht}
    \mathbf{H}_t = \mathbf{D}_t \mathbf{R}_t \mathbf{D}_t,
  \end{equation}
  but allows the correlation matrix \( \mathbf{R}_t \) to evolve over time. Correlations are derived from standardized residuals \( \boldsymbol{\varepsilon}_t = \mathbf{D}_t^{-1} Y_t \), through a dynamic process:
  \begin{equation}
    \label{eq:dcc_Qt}
    \mathbf{Q}_t = (1 - a - b)\bar{\mathbf{Q}} + a\, \boldsymbol{\varepsilon}_{t-1} \boldsymbol{\varepsilon}_{t-1}^\top + b\, \mathbf{Q}_{t-1},
  \end{equation}
  where \( \bar{\mathbf{Q}} \) is the unconditional covariance of \( \boldsymbol{\varepsilon}_t \), and \( a + b < 1 \) ensures the stationarity. The matrix \( \mathbf{Q}_t \) is then rescaled to yield the time-varying correlation matrix:
  \begin{equation}
    \label{eq:dcc_Rt}
    \mathbf{R}_t = \text{diag}(\mathbf{Q}_t)^{-1/2} \mathbf{Q}_t \text{diag}(\mathbf{Q}_t)^{-1/2}.
  \end{equation}

  By introducing dynamic correlations while preserving a parsimonious structure, the DCC model offers greater flexibility in modeling multivariate volatility dynamics. However, as the number of assets increases, the estimation of the correlation matrices becomes more complex. To circumvent this, the GO-GARCH model adopts a factor-based approach, as outlined below.

 \subsubsection{The Generalized Orthogonal GARCH Model}
\label{subsubsec422}

  The Generalized Orthogonal GARCH (GO-GARCH) model, introduced by \citet{vanderWeide2002}, extends the Orthogonal GARCH framework by allowing the mixing matrix to be any invertible matrix, thus relaxing the orthogonality constraint. The $r$-dimensional zero-mean return vector ${Y}_{t}$ is expressed as a linear combination of uncorrelated latent factors ${f}_{t}$:
  \begin{equation}
    {Y}_{t} = \mathbf{Z} {f}_{t}, \qquad {f}_{t} \sim \mathcal{N}(\mathbf{0}, \mathbf{H}_{t}),
  \end{equation}
  where $\mathbf{Z} \in \mathbb{R}^{r \times r}$ is an invertible matrix and $\mathbf{H}_{t} = \text{diag}(h_{1t}, \dots, h_{rt})$ is diagonal with each element following a univariate GARCH$(1,1)$ process:
  \begin{equation}
    h_{it} = \omega_{i} + \alpha_{i} f_{i,t-1}^{2} + \beta_{i} h_{i,t-1}, \quad \text{for } i = 1, \dots, r.
  \end{equation}

  The implied conditional covariance matrix of ${Y}_{t}$ is
  \begin{equation}
    \label{eq:gogarch_ht}
    \mathbf{\Sigma}_{t} = \mathbf{Z} \mathbf{H}_{t} \mathbf{Z}^{\top}.
  \end{equation}

   The estimation proceeds in two steps. First, a spectral decomposition of the unconditional covariance matrix $\mathbf{H} = \mathbb{E}[{Y}_{t} {Y}_{t}^{\top}]$ provides initial estimates of the eigenvectors $\mathbf{P}$ and the eigenvalues $\mathbf{A}$. Second, a rotation matrix $\mathbf{U}$ is estimated via the maximum likelihood method, yielding:
  \begin{equation}
    \mathbf{Z} = \mathbf{P} \mathbf{A}^{1/2} \mathbf{U}.
  \end{equation}

  The time-varying conditional correlation matrix is then recovered as
  \begin{equation} \label{eq:gogarch_rt}
    \mathbf{R}_{t} = \mathbf{D}_{t}^{-1} \mathbf{\Sigma}_{t} \mathbf{D}_{t}^{-1}, \qquad \mathbf{D}_t = \left(\mathbf{\Sigma}_t \odot \mathbf{I}\right)^{1/2},
  \end{equation}
 where $\odot$ denotes the Hadamard product.

  The GO-GARCH model is particularly suitable for high-dimensional settings. By factorizing returns into latent components, it reduces the parameter space and avoids the direct estimation of full conditional covariance matrices. Each latent process follows a simple GARCH structure, making the model both flexible and parsimonious in capturing complex volatility dynamics, as in our network-based analysis of OPEC countries' oil return volatilities.

  \section{Network-based Volatility models}
  \label{sec5}
  A network is defined as \( \mathcal{G} = (V, E) \), where \( V \) denotes the set of vertices (e.g., OPEC countries) and \( E \) the set of weighted edges capturing financial, geographic, or structural interdependencies; this structure underpins network-based log-ARCH models that embed both temporal and cross-sectional volatility dynamics to analyze and forecast spillovers.

  \subsection{Network log-ARCH models}
  \label{subsec51}

  The network log-ARCH model is an extension of spatiotemporal ARCH models \citep[see]
  []{ottoGeneralizedSpatialSpatiotemporal2018,ottoDynamicSpatiotemporalARCH2024},
  specifically designed to handle multivariate time series data in the context of
  network structures \citep{matteraNetworkLogARCHModels2024}. It captures both temporal and contemporaneous interactions
  in the volatility process, similar to the ARCH model, but adapted to a network setting. Unlike the previously reviewed univariate and multivariate GARCH models, these network-based approaches allow for volatility spillovers that occur instantaneously across the network in a GARCH-like manner. That is, the conditional variance in \eqref{eq:VarHt} also depends on adjacent nodes in \(\mathbf{Y}_t\) (e.g., $v=t$), instead of strictly conditioning on all $v<t$. Moreover, multivariate GARCH models are typically not suitable for a large number of cross-sections. This is a significant extension,
  as it implies that the current volatility is directly influenced by the volatilities of the neighboring nodes.

  
  In a network log-ARCH model, the process is observed at the nodes $V$, with the
  edges $E$ define the dependence structure in the volatility dynamics. This
  implies that the volatility at a given node is influenced not only by its own
  past observations but also by the volatility of its adjacent nodes at the same
  time point. Formally, the model builds upon the framework proposed by \citet{ottoDynamicSpatiotemporalARCH2024}
  for dynamic spatiotemporal ARCH models. The observed process $Y_{t}(s_{i})$ at
  time $t$ and node $v_{i}$ is given by
  \begin{equation}
    Y_{t}(v_{i}) = \sqrt{h_{t}(v_{i})}\varepsilon_{t}(v_{i}),
  \end{equation}
  where $h_{t}(v_{i})$ represents the volatility, and $\varepsilon_{t}(v_{i})$
  is a white-noise process. In the network log-ARCH model, the volatility
  $h_{t}(v_{i})$ depends not only on past values of $Y_{t-1}(v_{i})$ but also on
  the contemporaneous values from adjacent nodes, $\{Y_{t}(v_{j}) : j \in E_{i}\}$,
  where $E_{i}$ is the set of edges connected to node $v_{i}$.

  Let
  $\boldsymbol{h}^{*}_{t}= (\ln h^{2}_{t}(v_{1}), \ldots, \ln h^{2}_{t}(v_{n}))'$
  and
  $\boldsymbol{Y}^{*}_{t}= (\ln Y^{2}_{t}(v_{1}), \ldots, \ln Y^{2}_{t}( v_{n}))'$.
  The network log-ARCH process of order one can then be expressed as
  \begin{equation}
    \boldsymbol{h}^{*}_{t}= \boldsymbol{\omega}+ \mathbf{\Gamma}\boldsymbol{Y}^{*}
    _{t-1}+ \rho \mathbf{W}\boldsymbol{Y}^{*}_{t},
  \end{equation}
  where $\mathbf{W}$ is the matrix of edge weights (adjacency matrix) defining the
  degree of volatility spillovers across the network, $\rho$ is a parameter
  capturing the strength of these contemporaneous network interactions, and $\mathbf{\Gamma}$
  is a diagonal matrix of node-specific ARCH effects\footnote{Note that this is
  a specific extension proposed by \cite{matteraNetworkLogARCHModels2024}, deviating
  from the original spatiotemporal log-ARCH model of \cite{ottoDynamicSpatiotemporalARCH2024}.}.
  The term $\boldsymbol{\omega}$ is constant volatility intercepts. The matrix
  $\mathbf{W}$, analogous to a spatial weight matrix, defines how the volatility
  spillovers propagate through the network.

  The model can also be represented in an ARMA-like form of the log-squared observation:
  \begin{equation}
    \boldsymbol{Y}^{*}_{t}= \boldsymbol{\phi}_{0}+ \rho \mathbf{W}\boldsymbol{Y}^{*}
    _{t}+ \mathbf{\Gamma}\boldsymbol{Y}^{*}_{t-1}+ \boldsymbol{u}_{t},
  \end{equation}
  where $\boldsymbol{u}_{t}$ are the log-squared errors, and
  $\boldsymbol{\phi}_{0}$ is a constant vector. This formulation highlights the endogenous
  nature of the model, as $\boldsymbol{Y}^{*}_{t} = (\ln(Y^2_t(v_i)))_{i \in V}$ appears on both sides of the
  equation, necessitating specialized estimation techniques such as the Generalized
  Method of Moments (GMM), as proposed by
  \cite{ottoDynamicSpatiotemporalARCH2024} or quasi-maximum-likelihood
  estimators (QMLE) by \cite{otto2023dynamic} to effectively handle the endogeneity issue.

  \subsection{Choice of the Network Weight Matrices and Dissimilarity Measures}
  \label{subsec52} 
  
  The weight matrices \( \mathbf{W} \) are constructed based on the similarity between the log returns of the six OPEC countries and model-implied conditional correlations, measured using temporal distance metrics. We consider six alternative dissimilarity measures: Euclidean distances, correlation-based distances, Piccolo distances. These three dissimilarity measures are adopted from \citet{matteraNetworkLogARCHModels2024}. And we introduce three new conditional correaltion distances derived from the CCC-GARCH, DCC-GARCH, and GO-GARCH models, following the time series clustering methodologies outlined in \citet{Maharaj2019}.

  \textbf{(1) Euclidean Distance.} A raw-data-based approach computing pairwise
  distances between log-return observations:
  \begin{equation}
    \label{eq:euc_distance}
    d_{ij}^{\text{(Euc)}}= \sqrt{\sum_{t=1}^{T}\left( y_{t}^{(i)} - y_{t}^{(j)}\right)^{2}},
  \end{equation}
  where \( y_{t}^{(i)} \) denotes the oil price log return of country \( i \) at time \( t \).  
  The corresponding network weight is defined as:
  \begin{equation}
    W_{ij}^{\text{(Euc)}} = \frac{1}{d_{ij}^{\text{(Euc)}}} , \quad W_{ii}^{\text{(Euc)}} = 0,
  \end{equation}
  Therefore, smaller distances imply stronger connectivity.

  \textbf{(2) Correlation-Based Distance.} This feature-based approach uses the Pearson correlation coefficient \( \rho_{ij}^{\text{(corr)}} \) between countries \( i \) and \( j \), following the methodology of \citet{Mantegna1999}. The corresponding distance metric is:
  \begin{equation}
    \label{eq:corr_distance}
    d_{ij}^{\text{(corr)}} = \sqrt{2(1 - \rho_{ij}^{\text{(corr)}})},
  \end{equation}
  where \( \rho_{ij}^{\text{(corr)}} \) is computed from the log-return series \( \{ y_t^{(i)} \}_{t=1}^{T} \) as:
  \begin{equation}
    \label{eq:rho_definition}
    \rho_{ij}^{\text{(corr)}} = \frac{
      \frac{1}{T}  \displaystyle\sum_{t=1}^{T} y_{t}^{(i)} y_{t}^{(j)} - \left( \frac{1}{T} \sum_{t=1}^{T} y_{t}^{(i)} \right)
      \left( \frac{1}{T} \sum_{t=1}^{T} y_{t}^{(j)} \right)
    }{
      \sqrt{
        \left( \frac{1}{T} \displaystyle\sum_{t=1}^{T} (y_{t}^{(i)})^{2} - \left( \frac{1}{T} \sum_{t=1}^{T} y_{t}^{(i)} \right)^{2} \right)
        \left( \frac{1}{T} \displaystyle\sum_{t=1}^{T} (y_{t}^{(j)})^{2} - \left( \frac{1}{T} \sum_{t=1}^{T} y_{t}^{(j)} \right)^{2} \right)
      }
    }.
  \end{equation}

  This transformation maps high correlations (\( \rho_{ij}^{\text{(corr)}} \approx 1 \)) to small distances (\( d_{ij}^{\text{(corr)}} \approx 0 \)), providing a metric representation of similarity. The corresponding weight matrix is constructed as the inverse of the distance:
  \begin{equation}
    W_{ij}^{\text{(Corr)}} = \frac{1}{d_{ij}^{\text{(corr)}}}, \qquad W_{ii}^{\text{(Corr)}} = 0,
  \end{equation}
  so that stronger co-movements translate into higher weights in the network structure.

  \textbf{(3) Piccolo-Based Distance.} Following \citet{Piccolo1990} and \citet{matteraNetworkLogARCHModels2024}, model-based dissimilarities in volatility dynamics are computed using the log-ARCH coefficient vectors from:
  \begin{equation}
    \ln h_{t}^{(i)} = \omega_{i} + \sum_{p=1}^{P_i} \gamma_{ip} \ln (y_{t-p}^{(i)})^{2},
  \end{equation}
  where \( \gamma_{ip} \) are the autoregressive coefficients for country \( i \). Coefficient vectors are zero-padded for comparability when \( P_i \neq P_j \). This corresponds to a truncated AR($\infty$) representation in the sense of \citet{Piccolo1990}. The Piccolo distance is defined as:
  \begin{equation}
    \label{eq:logarch_distance}
    d_{ij}^{\text{(Pic)}} = \sqrt{\sum_{p=1}^{P}\left( \gamma_{ip} - \gamma_{jp} \right)^{2}},
  \end{equation}
  with the associated weight matrix:
  \begin{equation}
    W_{ij}^{\text{(Pic)}} = \frac{1}{d_{ij}^{\text{(Pic)}}}, \quad W_{ii}^{\text{(Pic)}} = 0.
  \end{equation}

  \textbf{(4) CCC-GARCH Distance.} Constant conditional correlations \( \rho_{ij}^{\text{(CCC)}} \), extracted from the matrix \( \mathbf{R} \) in \eqref{eq:ccc_Ht} as introduced by \citet{Bollerslev1990}, are used to construct a dissimilarity measure:
  \begin{equation}
    \label{eq:ccc_distance}
    d_{ij}^{\text{(CCC)}} = \sqrt{2(1 - \rho_{ij}^{\text{(CCC)}})}.
  \end{equation}
 
  To derive bounded and numerically stable network weights, we applied a nonlinear transformation of the squared distance:
  \begin{equation}
    \label{eq:ccc_weight}
    W_{ij}^{\text{(CCC)}} = \frac{1}{2(1 - \rho_{ij}^{\text{(CCC)}}) + 1}, \quad W_{ii}^{\text{(CCC)}} = 0.
  \end{equation}
 
  This formulation ensures finite weights and prevents singularities when \( \rho_{ij}^{\text{(CCC)}} \approx 1 \), which corresponds to a near-perfect correlation.

  \textbf{(5) DCC-GARCH Distance.} Based on the dynamic conditional correlations \( \rho_{ij,t}^{\text{(DCC)}} \), which are elements of the time-varying matrix \( \mathbf{R}_t \) in \eqref{eq:dcc_Rt} \citep{Engle2002}, we compute time-averaged correlations:
  \begin{equation}
    \label{eq:dcc_distance}
    {\rho}_{ij}^{\text{(DCC)}} = \frac{1}{T} \sum_{t=1}^{T} \rho_{ij,t}^{\text{(DCC)}}, \quad
    d_{ij}^{\text{(DCC)}} = \sqrt{2\left(1 - {\rho}_{ij}^{\text{(DCC)}}\right)}.
  \end{equation}
 
  The corresponding network weights are defined as follows:
  \begin{equation}
    \label{eq:dcc_weight}
    W_{ij}^{\text{(DCC)}} = \frac{1}{2\left(1 - {\rho}_{ij}^{\text{(DCC)}}\right) + 1}, \quad W_{ii}^{\text{(DCC)}} = 0.
  \end{equation}

  \textbf{(6) GO-GARCH Distance.} Based on the time-varying correlations \( \rho_{ij,t}^{\text{(GO)}} \) extracted from the matrix \( \mathbf{R}_t \) in \eqref{eq:gogarch_rt}, as proposed by \citet{vanderWeide2002}, we compute time-averaged correlations:
  \begin{equation}
    \label{eq:gogarch_avg_corr}
    {\rho}_{ij}^{\text{(GO)}} = \frac{1}{T} \sum_{t=1}^{T} \rho_{ij,t}^{\text{(GO)}}, \quad
    d_{ij}^{\text{(GO)}} = \sqrt{2(1 - {\rho}_{ij}^{\text{(GO)}})}.
  \end{equation}
 
  The associated network weights are as follows:
  \begin{equation}
    \label{eq:gogarch_weight}
    W_{ij}^{\text{(GO)}} = \frac{1}{2(1 - {\rho}_{ij}^{\text{(GO)}}) + 1}, \quad W_{ii}^{\text{(GO)}} = 0.
  \end{equation}

  All raw spatial weight matrices—whether based on Euclidean, correlation, Piccolo-Based, or MGARCH-derived dissimilarities—are symmetric by construction, satisfying \( W_{ij} = W_{ji} \) and \( W_{ii} = 0 \), thereby ensuring well-defined pairwise comparisons. The GO-GARCH-based dissimilarities offer a complementary perspective to CCC and DCC specifications: while the latter directly model pairwise correlations, GO-GARCH captures time-varying dependence through latent factor structures. This factor-based approach enhances estimation stability in high-dimensional settings and yields smoother inter-country correlation dynamics for network construction.

  \section{Results and interpretation}
  \label{sec6}
  This section presents the study’s empirical findings in three stages. First, we report the estimation results of the univariate and multivariate GARCH models using the full sample, which form the basis for constructing the network weight matrices. Second, we analyze the structural patterns revealed by the network plots to provide insight into the interdependencies among oil-exporting countries. Finally, we evaluate out-of-sample forecasting performance using an in-sample training period of 450 out of 503 months. We assess predictive accuracy by comparing Root Mean Squared Forecast Errors (RMSFE) and Mean Absolute Forecast Errors (MAFE), and by employing both standard and heteroscedasticity-robust Diebold-Mariano test statistics. We also apply the Model Confidence Set (MCS) procedure to rank the models from best to worst. The section concludes by discussing the advantages of Network Volatility models over standard MGARCH specifications. Furthermore, a full forecasting sensitivity analysis, assessing model performance across different in-sample training ranges, is provided in Table~\ref{tab_app:accuracy_sensitivity}.

  \subsection{Model Estimation}
  \label{sec61}
  
  As a foundation for the multivariate GARCH models employed in this study, we first estimate univariate GARCH(1,1) models assuming Student’s \textbf{t}-distributed innovations using the full dataset for this part, with estimated degrees of freedom capturing moderate to strong tail heaviness. This distributional choice is motivated by the non-Gaussian properties of monthly oil price log-returns, as evidenced by the leptokurtic density estimates and heavy-tailed Q-Q plots presented in Appendix Figure~\ref{fig_app:density_plots} and Figure~\ref{fig_app:qq_plots}.

  The univariate GARCH(1,1) estimates in Appendix Table~\ref{tab_app:garch_results} provide three key insights that motivate the transition to multivariate volatility modeling. First, all countries exhibit persistent volatility dynamics, with $\alpha_1 + \beta_1 \approx 1$, indicating long-lasting shock effects and stationary but highly persistent conditional variances. Second, the heterogeneity in the ARCH coefficients ($\alpha_1$) suggests differing short-run volatility responses, likely due to structural or informational asymmetries across markets. Third, the high log-likelihood values confirm the model's adequacy in capturing volatility clustering and time-varying second moments. The estimated degrees of freedom for the Student’s \textit{t}-distributed innovations further support the presence of heavy-tailed return distributions. Together, these findings justify the transition to multivariate GARCH models on three empirical grounds: (1) significant $\alpha_1$ and $\beta_1$ values confirm volatility clustering; (2) cross-country variation in parameters challenges the assumption of constant correlations; and (3) stationary yet persistent volatility ensures stable inputs for modeling interdependence. Hence, the univariate results offer both statistical and substantive foundations for estimating the CCC, DCC, and GO-GARCH models to examine cross-market volatility spillovers.

\begin{table}[!t]
	\centering
	\setlength{\tabcolsep}{4pt}  
	\caption{Estimation Results: CCC-GARCH and DCC-GARCH with Multivariate t-Distribution}
	\label{tab:cc_garch}
	\begin{tabular}{lcccccc}
		\toprule
		\textbf{Parameter} & \textbf{Algeria} & \textbf{Iran} & \textbf{Libya} & \textbf{Nigeria} & \textbf{Saudi Arabia} & \textbf{UAE} \\
		\midrule
		\multicolumn{7}{l}{\textit{Mean Equation}} \\
		$\mu$ & 0.0025 & 0.0058 & 0.0027 & 0.0016 & 0.0046 & 0.0045 \\
		& (0.5332) & (0.0931) & (0.5561) & (0.7898) & (0.7114) & (0.6237) \\
		\midrule
		\multicolumn{7}{l}{\textit{GARCH Equation}} \\
		$\omega$ & 0.0008 & 0.0013 & 0.0007 & 0.0006 & 0.0008 & 0.0007 \\
		& (0.2179) & (0.1332) & (0.4159) & (0.6511) & (0.8341) & (0.7862) \\
		$\alpha$ & 0.4210 & 0.5556 & 0.4335 & 0.4389 & 0.5191 & 0.4882 \\
		& (0.0000) & (0.0000) & (0.0000) & (0.0002) & (0.0520) & (0.0009) \\
		$\beta$ & 0.5549 & 0.4161 & 0.5655 & 0.5601 & 0.4799 & 0.5108 \\
		& (0.0000) & (0.0030) & (0.0000) & (0.0019) & (0.4477) & (0.3071) \\
		\textit{Shape ($\nu$)} & 8.7223 & 5.6097 & 9.0256 & 9.3528 & 5.2138 & 5.3147 \\
		& (0.0047) & (0.0000) & (0.0055) & (0.0063) & (0.0922) & (0.0041) \\
		\midrule
		\multicolumn{7}{l}{\textit{DCC Parameters (for the DCC-GARCH)}} \\
		$dcca_1$ & \multicolumn{6}{c}{0.0920 (0.0000)} \\
		$dccb_1$ & \multicolumn{6}{c}{0.8455 (0.0000)} \\
		\midrule
		\multicolumn{7}{l}{\textit{Model Fit}} \\
		Log-Likelihood & \multicolumn{6}{c}{7879.47 (DCC) \quad / \quad 7750.24 (CCC)} \\
		AIC & \multicolumn{6}{c}{-31.139 (DCC) \quad / \quad -30.633 (CCC)} \\
		BIC & \multicolumn{6}{c}{-30.736 (DCC) \quad / \quad -30.247 (CCC)} \\
		\bottomrule
	\end{tabular}
	\vspace{2mm}
	\begin{minipage}{\linewidth}
		\footnotesize \centering
		\textbf{Notes:} p-values are reported in parentheses below each coefficient estimate.
	\end{minipage}
\end{table}

   The estimation results for the CCC-GARCH and DCC-GARCH models, assuming a multivariate Student’s t-distribution, are presented in Table~\ref{tab:cc_garch} for the monthly log returns of the six OPEC countries. The unconditional means are mostly positive but statistically insignificant, indicating no persistent abnormal returns. In the variance equations, both the ARCH ($\alpha$) and GARCH ($\beta$) coefficients are significant for most countries, confirming short- and long-term volatility persistence, except for Saudi Arabia and UAE, where only ARCH coefficients are significant. The condition $\alpha + \beta < 1$ holds across countries, ensuring covariance stationarity. The estimated degrees of freedom, ranging between 5 and 9, further support the appropriateness of modeling innovations using a heavy-tailed multivariate Student’s \textit{t}-distribution. For correlation dynamics, the DCC parameters ($dcca_1 = 0.1016$, $dccb_1 = 0.8207$) are significant and satisfy $dcca_1 + dccb_1 < 1$, indicating stable and persistent correlations. Compared to CCC-GARCH, the DCC-GARCH model yields higher log-likelihood and lower AIC/BIC values, supporting its superior fit in capturing the time-varying joint volatility structure across the OPEC markets.

\begin{table}[!t]
	\centering
	\setlength{\tabcolsep}{8pt}  
	\caption{Estimation Results: GO-GARCH Model}
	\label{tab:go_garch}
	
	\begin{tabular}{lcccccc}
		\toprule
		& \multicolumn{6}{c}{\textbf{Latent Factors (F1–F6)}} \\
		\cmidrule(lr){2-7}
		\textbf{Parameter} & \textbf{F1} & \textbf{F2} & \textbf{F3} & \textbf{F4} & \textbf{F5} & \textbf{F6} \\
		\midrule
		
		\multicolumn{7}{l}{\textit{Rotation Matrix \( U \)}} \\
		\cmidrule(lr){1-7}
		Algeria          & -0.4404 & 0.1440 & -0.3351 & -0.3607 & 0.5218 & -0.5201 \\
		Iran             & 0.2449 & -0.1160 & -0.0040 & -0.6230 & -0.6064 & -0.4132 \\
		Libya            & 0.1822 & -0.2720 & 0.3498 & -0.6121 & 0.4591 & 0.4299 \\
		Nigeria          & 0.5941 & -0.2240 & 0.2464 & 0.2940 & 0.3631 & -0.5637 \\
		Saudi Arabia     & 0.5994 & 0.4260 & -0.6178 & -0.1122 & 0.1315 & 0.2183 \\
		UAE              & 0.0239 & 0.8120 & 0.5683 & -0.0898 & 0.0063 & -0.0925 \\
		\midrule
		
		\multicolumn{7}{l}{\textit{GARCH Parameters}} \\
		\cmidrule(lr){1-7}
		\(\omega\)   & 0.1233 & 0.0771 & 0.1012 & 0.1516 & 0.1329 & 0.1312 \\
		\(\alpha_1\) & 0.2882 & 0.4033 & 0.2560 & 0.2796 & 0.8065 & 0.5505 \\
		\(\beta_1\)  & 0.6021 & 0.5937 & 0.6651 & 0.5884 & 0.1925 & 0.4316 \\
		\midrule
		
		\multicolumn{7}{l}{\textit{Model Fit}} \\
		\cmidrule(lr){1-7}
		Log-Likelihood & \multicolumn{6}{c}{7748.56}  \\
		AIC            & \multicolumn{6}{c}{-30.74}  \\
		BIC            & \multicolumn{6}{c}{-30.59}  \\
		\bottomrule
	\end{tabular}
	
	\vspace{1ex}
	\begin{minipage}{0.88\textwidth}
		\footnotesize
		\textbf{Notes:} The GO-GARCH model estimates latent factors (F1–F6) with constant mean and GARCH(1,1) volatility dynamics. The mixing matrix \( A \) was omitted for brevity.
	\end{minipage}
\end{table}

  The GO-GARCH model offers a structural decomposition of volatility across OPEC countries by identifying the latent sources of co-movement. It supports a more parsimonious representation by modeling each latent factor with a univariate GARCH(1,1) process, thereby reducing dimensionality while preserving essential volatility dynamics. As shown in Table~\ref{tab:go_garch}, the estimated rotation matrix \( U \) reveals heterogeneous exposures of countries to different latent factors—some associated with short-term shocks and others with more persistent volatility—consistent with the covariance stationarity condition. This setup contrasts with the CCC and DCC models, which directly model the pairwise correlations. While DCC-GARCH flexibly captures time-varying dependence structures suitable for dynamic adjacency matrices, GO-GARCH provides stable and interpretable latent channels that inform the network’s underlying topology. Overall, GO-GARCH complements the DCC by enriching the network modeling framework with structural insights into systemic volatility transmission.

  \subsection{Network Construction}
  \label{sec62}

  Following the estimation of the CCC-GARCH, DCC-GARCH, and GO-GARCH models using the full dataset, we constructed six distance matrices to support the network analysis. Two data-driven matrices were derived from monthly log returns using Euclidean and correlation-based dissimilarities. Subsequently, four model-based dissimilarity matrices were computed: one using the truncated AR($\infty$) representations of log-squared returns following \citet{Piccolo1990}, and three novel structures derived from the conditional correlation matrices produced by the CCC, DCC, and GO-GARCH models. All matrices were constructed following the methodology detailed in Subsection~\ref{subsec52}, ensuring comparability and coherence in the subsequent network-modeling framework.

  \begin{figure} [h!]
    \centering
    \includegraphics[width=0.8\linewidth]{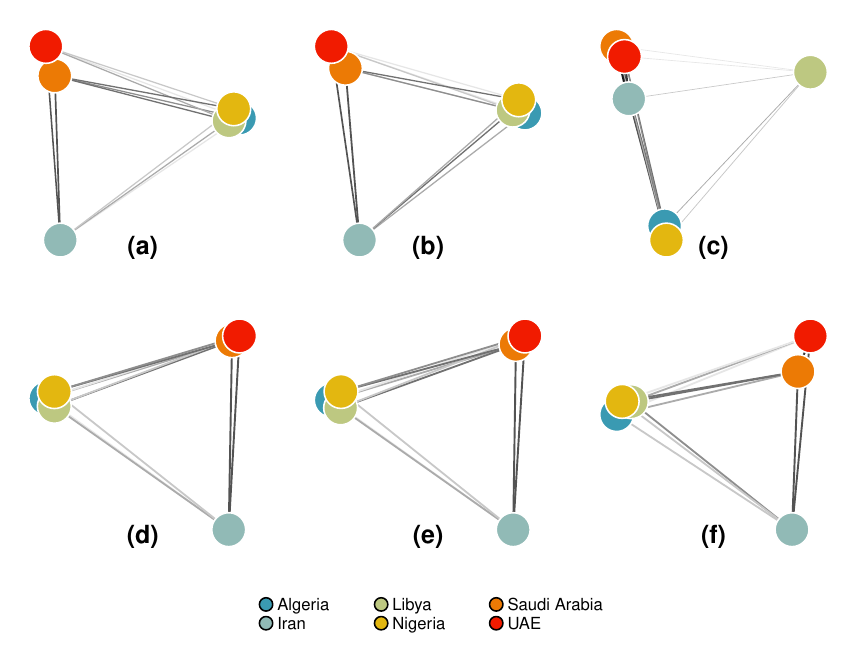}
    \caption{Network plots of the six OPEC oil-exporting countries based on six different distance measures. Raw data-based distances: (a) Euclidean distance, (b) Correlation-based distance. Model-based distances: (c) Piccolo-based distance, (d) CCC-GARCH, (e) DCC-GARCH, and (f) GO-GARCH. The grey colour scale of the edges is proportional to the weight degree of the connections.}
    \label{fig:network_plots}
  \end{figure}
  
  The network structures depicted in Figure~\ref{fig:network_plots} illustrate the oil price interdependencies among the six OPEC countries. These networks were derived solely from monthly log returns. Notably, this methodology yielded coherent and interpretable structures across all modeling approaches without relying on external variables or geospatial data, underscoring the robustness of the proposed clustering approaches.
  
  The data-driven networks—Euclidean (a) and correlation-based (b)—highlight regional proximity, with Algeria, Libya, and Nigeria forming a North African cluster, likely reflecting geographic and economic commonalities. In contrast, the model-based networks reveal volatility-driven connections. The Piccolo distance network (c), derived from \citet{Piccolo1990}’s AR($\infty$) structure, displays a distinct topology suggesting the influence of long-memory volatility dynamics. However, this approach yields a weaker and less representative network structure, characterized by sparse connectivity and minimal edge formation. It demonstrates limited evidence of spillover effects or interconnectedness in this context. Furthermore, this topology lacks robustness, as evidenced by the inconsistent network representations across different in-sample training periods shown in Appendix Figure~\ref{fig_app:six_weight_matrices}.
  
  Conversely, the CCC-GARCH (d), DCC-GARCH (e), and GO-GARCH (f) models provide the most accurate and consistent spatial representation of network formation compared to the heuristic measures, especially the GO-GARCH based weight matrix. These models reveal critical information about the connectivity between OPEC members, exhibiting a strong Middle Eastern core with Saudi Arabia and the UAE emerging as central hubs. Saudi Arabia, in particular, appears as the primary source of risk spillovers and the central node for the majority of OPEC members, highlighting its capacity to propagate systemic risk. This dominant role aligns with its status as the largest oil producer within the organization \citep{fattouhOPECPricingPower2007,Alkhathlan2014,AlRousan2018}. The analysis also reveals the nuanced position of Iran, whose shifting connections are influenced by external constraints—such as sanctions and production limits—that also impact its macroeconomic indicators \citep{Nakhli2021}.
  
  These networks, constructed solely from statistical features and dissimilarity measures, demonstrate how time-series clustering methods—especially the conditional correlation MGARCH-informed networks—effectively capture latent interdependencies and volatility transmissions. These interdependencies may also reflect broader macroeconomic shocks and geopolitical frictions, as emphasized by \citet{kilianEconomicEffectsEnergy2008}. Moreover, as noted by \citet{Mignon2024}, such factors are essential for understanding systemic risk and volatility propagation within the global oil market. These visualizations identify dominant clusters and key players in the OPEC cartel, reflecting real economic dependencies under varying model assumptions. They serve as the structural foundation for the spillover channels assessed in the forecasting performance evaluation in the next subsection. (We refer the reader to Appendix Figure~\ref{fig_app:six_weight_matrices} for a representation of the weight matrices under different initial training windows $T_0$ employed for the rolling window forecasting session).

  \subsection{Forecasting Results}
  \label{sec63}
  
  First, using the in sample period of $T_0= 450$ months (February 1983-August 2020), we estimate the network-based volatility dynamics using the Generalized Method of Moments (GMM) framework proposed by \citet{ottoDynamicSpatiotemporalARCH2024} for dynamic spatiotemporal ARCH models. The models incorporate six different network weight matrices estimated within the in-sample period for each model. Table~\ref{tab:estimation_results_450} presents the estimation results for the in-sample for each network approach. Across all cases, we observe a pronounced instantaneous network ARCH effect ($\rho$), while the country-specific temporal ARCH effects ($\gamma_1, \ldots, \gamma_6$) remain relatively small and partially statistically insignificant.

\begin{table}[h!]
	\centering
	\caption{GMM Estimation Results of the Dynamic Spatiotemporal ARCH Model (In-Sample $T=450$)}
	\label{tab:estimation_results_450}
	\setlength{\tabcolsep}{4pt} 
	\begin{tabular}{lrrrrrrrr}
		\toprule
		\textbf{Network Weight Matrix} & \textbf{$\rho$} & \textbf{$\gamma_1$} & \textbf{$\gamma_2$} & \textbf{$\gamma_3$} & \textbf{$\gamma_4$} & \textbf{$\gamma_5$} & \textbf{$\gamma_6$} & \textbf{$\sigma^2$} \\
		\midrule
		\textbf{Euclidean-based} & 0.948 & -0.035 & 0.059 & -0.053 & -0.105 & 0.226 & 0.220 & 1.553 \\
		Std. Error             & 0.058 & 0.048 & 0.042 & 0.052 & 0.057 & 0.043 & 0.049 & \\
		t-statistic            & 16.374 & -0.732 & 1.410 & -1.018 & -1.840 & 5.223 & 4.528 & \\
		\midrule
		\textbf{Correlation-based} & 0.951 & -0.031 & 0.050 & -0.057 & -0.103 & 0.229 & 0.216 & 1.540 \\
		Std. Error                & 0.058 & 0.048 & 0.042 & 0.052 & 0.057 & 0.043 & 0.048 & \\
		t-statistic               & 16.461 & -0.650 & 1.202 & -1.090 & -1.809 & 5.318 & 4.466 & \\
		\midrule
		\textbf{Piccolo-based} & 1.232 & -0.145 & 0.115 & -0.144 & -0.066 & 0.344 & 0.362 & 2.503 \\
		Std. Error             & 0.105 & 0.065 & 0.053 & 0.070 & 0.074 & 0.054 & 0.058 & \\
		t-statistic            & 11.726 & -2.216 & 2.187 & -2.059 & -0.900 & 6.373 & 6.223 & \\
		\midrule
		\textbf{CCC-GARCH-Based}            & 0.923 & -0.033 & 0.044 & -0.041 & -0.073 & 0.202 & 0.122 & 1.635 \\
		Std. Error             & 0.056 & 0.051 & 0.045 & 0.054 & 0.058 & 0.045 & 0.051 & \\
		t-statistic            & 16.479 & -0.641 & 0.964 & -0.757 & -1.254 & 4.453 & 2.386 & \\
		\midrule
		\textbf{DCC-GARCH-Based}            & 0.924 & -0.032 & 0.044 & -0.041 & -0.073 & 0.201 & 0.123 & 1.635 \\
		Std. Error             & 0.056 & 0.051 & 0.045 & 0.054 & 0.058 & 0.045 & 0.051 & \\
		t-statistic            & 16.479 & -0.638 & 0.983 & -0.772 & -1.265 & 4.446 & 2.397 & \\
		\midrule
		\textbf{GO-GARCH-Based}       & 0.922 & -0.033 & 0.041 & -0.039 & -0.069 & 0.201 & 0.128 & 1.647 \\
		Std. Error             & 0.056 & 0.051 & 0.045 & 0.054 & 0.058 & 0.045 & 0.051 & \\
		t-statistic            & 16.411 & -0.652 & 0.897 & -0.733 & -1.182 & 4.412 & 2.484 & \\
		\bottomrule
	\end{tabular}
	
	\vspace{1ex}
	\begin{minipage}{0.95\textwidth}
		\footnotesize
		\textbf{Notes:} $\hat{\rho}$ is the spatial autoregressive coefficient; $\hat{\gamma}_j$ ($j = 1,\ldots,6$) are the conditional variance parameters for each of the six OPEC countries. The six weight matrices are based on: Euclidean, Correlation, Piccolo, CCC-GARCH, DCC-GARCH, and GO-GARCH distances.
	\end{minipage}
\end{table}

Following the estimation of the network-based volatility models, we assess their out-of-sample forecasting performance using two standard loss functions: root mean squared forecast error (RMSFE) and mean absolute forecast error (MAFE). These metrics are computed over the full forecast horizon $T$ and across all $N$ countries. Let $y_{i,t} \equiv Y_{t}^{*}(v_i) = \ln Y_{t}^{2}(v_i)$ denote the observed log-squared return and $\hat{y}_{i,t}$ its one-step-ahead forecast for country $i$ at time $t$. To evaluate the predictive accuracy of the full multivariate system, we compute global metrics aggregated across all countries and time periods:
\begin{align}
	\mathrm{RMSFE} &= \sqrt{\frac{1}{NT} \sum_{i=1}^{N} \sum_{t=1}^{T} (y_{i,t} - \hat{y}_{i,t})^2}, \\
	\mathrm{MAFE}  &= \frac{1}{NT} \sum_{i=1}^{N} \sum_{t=1}^{T} |y_{i,t} - \hat{y}_{i,t}|.
\end{align}
To account for sampling uncertainty, we compute 95\% confidence intervals for these metrics using a stationary bootstrap procedure with 2,000 replications.

To formally test for differences in predictive accuracy, we employ the Diebold--Mariano (DM) test \citep{Diebold1995, Diebold2002}. Let $e^{(m)}_{i,t} = y_{i,t} - \hat{y}^{(m)}_{i,t}$ denote the forecast error from model $m$ for country $i$ at time $t$. We define the pooled loss differential between a candidate model and the benchmark (Net. GO-GARCH) across all space-time observations as:
\begin{equation}
	d_{i,t} = L(e^{(\text{bench})}_{i,t}) - L(e^{(\text{cand})}_{i,t}), \qquad \text{with } L(e) = e^2 \text{ or } |e|,
	\label{eq:loss_diff}
\end{equation}
and apply the test to the pooled sequence $\{d_{i,t}\}_{i=1,\ldots,N; \, t=1,\ldots,T}$ containing all $NT$ observations. The standard DM test statistic is computed as:
\begin{equation}
	\mathrm{DM} = \frac{\bar{d}}{\sqrt{\widehat{\mathrm{Var}}(\bar{d})}}, \qquad 
	\bar{d} = \frac{1}{NT} \sum_{i=1}^{N} \sum_{t=1}^{T} d_{i,t},
	\label{eq:dm_stat}
\end{equation}
where $\widehat{\mathrm{Var}}(\bar{d})$ denotes the heteroskedasticity and autocorrelation consistent (HAC) variance estimator of \citet{Newey1987}, with truncation lag $k - 1 = 0$ for one-step-ahead forecasts.

Given the strong evidence of heteroskedasticity in our data—demonstrated by the distributional properties in Appendix Figures~\ref{fig_app:density_plots} and~\ref{fig_app:qq_plots}—and the time-varying volatility inherent in our standard MGARCH and Network Log-ARCH frameworks, the standard Diebold-Mariano test may suffer from size distortions or power loss. To address this, we employ the heteroskedasticity-robust test for equal forecast accuracy proposed by \citet{Harvey2024}. When applied to our pooled panel structure, the method treats the $NT$ loss differentials as a single sequence allowing for unconditional heteroskedasticity:
\begin{equation}
	d_{i,t} = c + \sigma(i,t) \epsilon_{i,t}, \quad \text{with } \mathrm{Var}(d_{i,t}) = \sigma^2(i,t),
	\label{eq:harvey_model}
\end{equation}
where $\sigma(\cdot)$ is a deterministic positive volatility function indexed over the pooled observations and $\{\epsilon_{i,t}\}$ is a mean-zero error process. The null hypothesis of equal forecast accuracy is $H_0: c = 0$. To test this, the method rescales the loss differential by a nonparametric estimate of the time-varying volatility, yielding the test statistic:
\begin{equation}
	\mathrm{DM}_{\text{het}} = \frac{\sqrt{NT} \cdot \overline{d^*}}{\sqrt{\widehat{\Omega}(d^*)}}, \quad \text{where} \quad d^*_{i,t} = \frac{d_{i,t}}{\hat{\sigma}_{i,t}},
	\label{eq:harvey_dm}
\end{equation}
and $\widehat{\Omega}(d^*)$ is a long-run variance estimator applied to the standardized series.

The spot volatility $\hat{\sigma}_{i,t}$ is estimated using a kernel smoother with a leave-$k$-out cross-validation scheme to handle serial correlation in the pooled sequence. Specifically, treating the $NT$ observations as indexed sequentially, the variance at each position is estimated as:
\begin{equation}
	\hat{\sigma}^2_{i,t} = \sum_{j} w_{j,(i,t)} d_{j}^2, \quad 
	w_{j,(i,t)} \propto K_h\left(\frac{\text{pos}(i,t)-\text{pos}(j)}{NT}\right) \cdot \mathbb{1}\{|\text{pos}(i,t)-\text{pos}(j)| \geq k\},
	\label{eq:spot_variance}
\end{equation}
where $K_h(\cdot)$ is a Gaussian kernel with bandwidth $h$, pos$(i,t)$ denotes the sequential position in the pooled series, and the indicator function ensures that the estimation excludes the $2k$ nearest neighbors. Following \citet{Harvey2024}, we set $k=20$. Under the null hypothesis and standard regularity conditions, the heteroskedasticity-adjusted statistic follows a standard normal distribution asymptotically:
\begin{equation}
	\mathrm{DM}_{\text{het}} \xrightarrow{d} N(0,1).
	\label{eq:harvey_asymptotic}
\end{equation}

This adjustment ensures that the test maintains correct size even under severe volatility clustering while achieving substantial power gains over the standard DM test when heteroskedasticity is present. We report both the standard and heteroskedasticity-adjusted results, computing two-sided $p$-values at the 5\% significance level.

 \subsubsection{Out-of-Sample Forecast Evaluation}
 \label{sec631}
 
\begin{table*}[!ht]
	\caption{Forecast Performance Evaluation: Bootstrap Confidence Intervals and Diebold-Mariano Tests (Benchmark: Net. GO-GARCH)}
	\label{tab:forecast_eval_main}
	\centering
	\renewcommand{\arraystretch}{1.2}
	\resizebox{0.95\textwidth}{!}{%
		\begin{tabular}{lcc|cc|cc}
			\toprule
			\multirow{2}{*}{\textbf{Model}} 
			& \multicolumn{2}{c|}{\textbf{RMSFE CI}} 
			& \multicolumn{2}{c|}{\textbf{MAFE CI}} 
			& \multicolumn{2}{c}{\textbf{DM $p$-values}} \\
			\cmidrule(lr){2-3} \cmidrule(lr){4-5} \cmidrule(lr){6-7}
			& Lower & Upper & Lower & Upper & Standard & \textbf{Robust (Harvey)} \\
			\midrule
			Std. GO-GARCH            & 5.7257 & 6.8233 & 5.4276 & 6.4027 & $<0.001^{***}$ & $<0.001^{***}$ \\
			Piccolo-Diss             & 3.9386 & 5.0313 & 3.5115 & 4.4813 & $<0.001^{***}$ & $<0.001^{***}$ \\
			Correlation-Diss         & 2.3045 & 3.2408 & 1.7465 & 2.5194 & $<0.001^{***}$ & $<0.001^{***}$ \\
			Euclidean-Diss           & 2.2994 & 3.2372 & 1.7418 & 2.5150 & $<0.001^{***}$ & $<0.001^{***}$ \\
			\midrule
			Std. CCC-GARCH           & 1.9486 & 3.0546 & 1.3617 & 2.2305 & $0.0442^{**}$  & $0.0447^{**}$  \\
			Std. DCC-GARCH           & 1.8846 & 2.8454 & 1.2840 & 2.0665 & $0.4704$       & $0.4803$       \\
			Net. DCC-GARCH           & 1.9456 & 2.6817 & 1.5248 & 2.1810 & $0.2060$       & $0.1110$       \\
			Net. CCC-GARCH           & 1.9455 & 2.6815 & 1.5246 & 2.1812 & $0.2167$       & $0.1071$       \\
			\textbf{Net. GO-GARCH}   & 1.9457 & 2.6802 & 1.5250 & 2.1835 & --- & --- \\
			\bottomrule
		\end{tabular}
	}
	\vspace{1.5ex}
	\begin{minipage}{0.95\textwidth}
		\footnotesize
		\textbf{Notes:} 95\% bootstrap confidence intervals (RMSFE/MAFE) are based on 2,000 replications. 
		The \textbf{Standard DM} column reports $p$-values from the \citet{Diebold2002} test assuming homoscedasticity. 
		The \textbf{Robust (Harvey)} column reports the $DM^*$ statistic from \citet{Harvey2024}, which corrects for heteroscedasticity in the loss differentials. 
		All tests are performed against the benchmark model, \textbf{Net. GO-GARCH}. 
		Significance levels: $^{***}p<0.001$, $^{**}p<0.05$.
	\end{minipage}
\end{table*}

The out-of-sample forecasting exercise is conducted using a rolling window framework with an initial estimation window of $T_0 = 450$ monthly observations, covering the period from February 1983 to July 2020. The remaining 53 observations, spanning August 2020 to December 2024, are reserved for forecast evaluation. This choice ensures a sufficiently large training sample while preserving a meaningful evaluation period.

Table~\ref{tab:forecast_eval_main} reports 95\% bootstrap confidence intervals for the Root Mean Squared Forecast Error (RMSFE) and Mean Absolute Forecast Error (MAFE) based on 2,000 replications. Point estimates for all models under the baseline specification, along with robustness checks across alternative training-window lengths and zero-adjustment schemes, are provided in Appendix Table~\ref{tab_app:accuracy_sensitivity}. The comparison focuses on standard multivariate GARCH specifications and a class of Network Log-ARCH models constructed using GARCH-informed distance matrices.

Several key findings emerge from the forecast evaluation. First, the Network Log-ARCH model based on GO-GARCH distances (Net. GO-GARCH) achieves the lowest RMSFE (2.33) among all network-based and standard mutivaraite GARCH specifications and serves as our benchmark model. While the Standard DCC-GARCH model attains a lower MAFE (1.67 vs.\ 1.85), the Net. GO-GARCH exhibits tighter confidence intervals and demonstrates superior performance on the RMSFE criterion, which penalizes large forecast errors more heavily. The benchmark's performance is uniformly superior to that of standard GO-GARCH and CCC-GARCH models, as well as network specifications based on simpler dissimilarity measures such as Euclidean, correlation-based, and Piccolo distances. This result underscores the importance of embedding economically meaningful dependence structures into the construction of network weights.

Second, under the baseline specification ($T_0=450$), the network-based models employing GARCH-informed distance matrices consistently outperform their standard multivariate counterparts on the RMSFE criterion. For instance, Net. GO-GARCH (2.33) drastically outperforms Std. GO-GARCH (6.27). As documented in Appendix Table~\ref{tab_app:accuracy_sensitivity}, this ranking remains robust across zero-adjustment perturbations. However, we note that the comparison with Standard DCC-GARCH is sensitive to the training window size; while the Network approach yields lower RMSFE in the extended sample ($T_0=450$), the Standard DCC model remains highly competitive in shorter windows (e.g., $T_0=200$). The close similarity in forecasting accuracy across the network variants (point estimates within 0.001 for RMSFE) suggests that the primary source of improvement lies in the spatial specification itself rather than the specific standard MGARCH model used to derive the weights.

Third, the formal statistical tests confirm the competitiveness of the proposed framework. Both the standard Diebold--Mariano test \citep{Diebold2002} and the heteroskedasticity-robust modification proposed by \citet{Harvey2024} fail to reject the null hypothesis of equal predictive accuracy between the Net. GO-GARCH benchmark and the Standard DCC-GARCH model at conventional significance levels ($p = 0.47$ and $p = 0.48$, respectively). This indicates that the Network Log-ARCH framework integratingating the CCC-, DCC- and GO-GARCH based weights achieves a level of accuracy statistically indistinguishable from the fully dynamic DCC model in the baseline sample.

From a methodological standpoint, this result is particularly informative. The network structure allows for the direct incorporation of instantaneous volatility spillovers through GARCH-like spatial terms, while estimation is carried out via GMM rather than high-dimensional likelihood maximization. Consequently, the network approach attains forecasting accuracy comparable to—and on the RMSFE criterion superior to—fully dynamic multivariate GARCH models, but with a substantially lower computational burden. This advantage is especially relevant in rolling-window forecasting environments and applications involving interconnected markets, such as OPEC oil returns.

\subsubsection{Model Confidence Set (MCS) Analysis}

To assess the statistical significance of the forecast rankings and control for data-snooping bias, we apply the Model Confidence Set (MCS) procedure of \citet{hansenModelConfidenceSet2011}. Unlike pairwise comparisons, the MCS allows us to identify a subset of models—the Superior Set of Models (SSM), denoted as $\widehat{\mathcal{M}}^*_{1-\alpha}$—that contains the best forecasting model with a given level of confidence ($1-\alpha$). We set $\alpha=0.05$.

The results are reported in Table~\ref{tab:mcs_results}, which presents the model rankings, the MCS $p$-values based on the Range ($T_R$) and Semi-Quadratic ($T_{SQ}$) statistics, and the average loss values.

\begin{table*}[!ht]
	\centering
	\caption{Model Confidence Set (MCS) Results ($\alpha=0.05$, $B=5000$)}
	\label{tab:mcs_results}
	\renewcommand{\arraystretch}{1.3}
	\setlength{\tabcolsep}{8pt}
	\footnotesize
	\begin{tabular*}{\textwidth}{@{\extracolsep{\fill}}lcccccc@{}}
		\toprule
		\textbf{Model} & \textbf{Rank} & $v_{T_{SQ}}$ & \textbf{MCS$_{T_{SQ}}$} & $v_{T_{R}}$ & \textbf{MCS$_{T_{R}}$} & \textbf{Loss (MSE)} \\
		\midrule
		\textbf{Net. GO-GARCH}   & \textbf{1} & \textbf{-0.6966} & \textbf{1.0000} & \textbf{-0.1922} & \textbf{1.0000} & \textbf{5.5578} \\
		Net. CCC-GARCH           & 2 & -0.6822 & 1.0000 & 1.3743 & 0.9996 & 5.5625 \\
		Net. DCC-GARCH           & 3 & -0.6805 & 1.0000 & 1.4463 & 0.4882 & 5.5631 \\
		Std. DCC-GARCH           & 4 & -0.1245 & 1.0000 & 0.1922 & 0.9996 & 5.7002 \\
		Std. CCC-GARCH           & 5 &  1.3185 & 0.2394 & 1.6410 & 0.3502 & 6.4114 \\
		\bottomrule
	\end{tabular*}
	\vspace{0.5ex}
	\begin{minipage}{\textwidth}
		\small
		\textbf{Note:} The table reports the models retained in the Superior Set $\widehat{\mathcal{M}}^*_{95\%}$. Four models (Euclidean-Diss, Correlation-Diss, Piccolo-Diss, and Standard GO-GARCH) were eliminated from the MCS due to significantly inferior performance ($p < 0.05$). The \textbf{Net. GO-GARCH} model achieves the lowest loss and holds the top rank. Columns $v$ and MCS report the test statistics and $p$-values for the Semi-Quadratic ($T_{SQ}$) and Range ($T_{R}$) tests, respectively.
	\end{minipage}
\end{table*}

The MCS outcomes strongly corroborate the conclusions drawn from the forecast analysis. As shown in Table~\ref{tab:mcs_results}, the Net. GO-GARCH model achieves the lowest average loss (5.5578) and occupies the top rank within the Superior Set. It is closely followed by the Net. CCC-GARCH and Net. DCC-GARCH specifications. Notably, all three GARCH-informed network models exhibit losses significantly lower than the Standard DCC-GARCH (5.7002), though the latter is still retained in the MCS with a high inclusion probability ($p=1.00$).

In contrast, the models based on simpler dissimilarity measures (Euclidean, correlation-based, and Piccolo distances), as well as the Standard GO-GARCH specification, are eliminated from the MCS. This exclusion indicates that their predictive performance is significantly inferior to the benchmark at the 5\% level. Although the Standard CCC-GARCH model is technically retained in the set, it exhibits the weakest performance among the survivors (Loss = 6.4114) and a markedly lower $p$-value ($0.2394$), suggesting it is a borderline case.

Overall, these findings provide compelling evidence that GARCH-informed network topologies capture cross-market dependence structures more effectively than both simpler distance-based networks and traditional multivariate GARCH models. A central contribution highlighted by these results is the efficiency of the Network Log-ARCH framework: the Net. GO-GARCH model not only outperforms the fully dynamic Standard DCC-GARCH in terms of pure accuracy (RMSFE), but it does so using a computationally efficient structure that relies on static time-averaged weight matrices rather than time-varying covariance estimation. This trade-off between forecasting accuracy and computational feasibility is particularly relevant for high-dimensional applications in interconnected energy or financial markets. Furthermore, this result significantly extends the findings of \citet{matteraNetworkLogARCHModels2024}. While their study demonstrated the superiority of the network framework over independent univariate log-ARCH variants, we provide novel evidence that—when constructed with GARCH-informed weights—the Network Log-ARCH model rivals, and in key metrics surpasses, fully dynamic multivariate GARCH specifications.

\subsection{Computational Tractability and Methodological Contributions}
\label{sec64}
The primary challenge in multivariate volatility modeling is the curse of dimensionality, where traditional specifications like the full BEKK-GARCH \citep{Engle1995} or VECH-GARCH suffer from parameter proliferation, and dynamic conditional correlation (DCC) models face extreme computational bottlenecks due to repeated matrix inversions and iterative likelihood optimization. To resolve this, the Network Log-ARCH framework \citep{matteraNetworkLogARCHModels2024} parameterizes cross-sectional dependence through a spatial weight matrix and employs a closed-form Generalized Method of Moments (GMM) estimator to compute a minimal set of parameters entirely independent of the asset dimension $N$. As detailed in Table~\ref{tab_app:computational_comparison}, this theoretical tractability translates into massive empirical gains: the network-augmented models operate approximately 27,000 to 62,000 times faster than standard MGARCH models, utilize roughly 51\% less peak memory, and exhibit negligible computational sensitivity to expanding rolling window sizes ($T_0=200$ to $450$). Crucially, this massive computational advantage does not come at the cost of forecasting accuracy, as long as the network structure reflects true market dependencies. By replacing the heuristic clustering metrics frequently utilized in spatial econometrics \citep[e.g.,][]{matteraNetworkLogARCHModels2024} with topologies derived from the conditional covariances of CCC-, DCC-, and GO-GARCH models, the network framework successfully captures latent time-varying dependencies. Consequently, these GARCH-informed network specifications produce forecasts statistically indistinguishable from the fully dynamic standard DCC-GARCH model---confirmed by pairwise Diebold--Mariano tests and joint retention in the Model Confidence Set (MCS) at the 5\% level. Conversely, simpler dissimilarity measures are strictly excluded from the MCS, including the topologically unstable Piccolo distance, which fails to capture robust structural linkages within the OPEC cartel as evidenced by Figures~\ref{fig:network_plots} and~\ref{fig_app:six_weight_matrices}. 

By matching the predictive accuracy of the DCC-GARCH benchmark at a fraction of the computational cost, the GARCH-informed Network volatility specifications lies on the absolute efficiency frontier of volatility models. This dual advantage directly enables real-time risk management under strict latency constraints and allows for the robust modeling of massive, highly interconnected global financial portfolios where standard specifications would otherwise fail to converge.

  \section{Conclusion}
  \label{sec7}
  
  This study introduces a computationally efficient framework for modeling volatility spillovers in interconnected markets by combining spatial econometric techniques in a network framewrok for volatility modeling with multivariate GARCH insights. We propose Network Log-ARCH models that parameterize cross-sectional dependence through GARCH-informed weight matrices, offering a scalable alternative to traditional multivariate GARCH specifications that suffer from the curse of dimensionality in high-dimensional settings.
  
  Applied to a selected panel of oil-exporting OPEC countries, our empirical analysis yields three key findings. First, network models constructed from GARCH-derived distance matrices (CCC, DCC, GO-GARCH) substantially outperform simpler network topologies based on Euclidean, correlation, or Piccolo-based distances, underscoring the importance of embedding economically meaningful dependence structures into the network construction. Second, the proposed Network log-ARCH models that embed these MGARCH-informed weight matrices achieve the lowest RMSFE. Their forecast accuracy is statistically indistinguishable from the fully dynamic standard DCC-GARCH model, as confirmed by both pairwise Diebold--Mariano tests and Model Confidence Set procedures, yet they require substantially fewer parameters and avoid iterative covariance matrix estimation. Third, despite relying on time-averaged static weight matrices, the GARCH-informed network models effectively capture persistent volatility linkages and contemporaneous spillover effects across OPEC members, proving that the conditional correlation structures extracted during the training period remain highly robust over the forecast horizon and across different in-sample window sizes.
  
  From a methodological perspective, the Network Log-ARCH framework resolves the trade-off between model complexity and computational tractability. By estimating only a small set of parameters via GMM, independent of the system dimension $N$, the approach scales efficiently to high-dimensional portfolios where standard BEKK or full DCC specifications become practically infeasible. This computational advantage is particularly valuable in rolling-window forecasting environments and real-time risk management applications.
  
  While the current framework relies on static time-averaged network topologies, our results suggest a natural extension: incorporating time-varying weight matrices $W_t$ to capture evolving volatility transmission channels within a fully dynamic Network log-ARCH specification. Such dynamic network structures could further enhance forecasting performance during periods of structural change or market stress, particularly in volatile energy markets where geopolitical events and supply disruptions induce regime shifts. Future research could also expand the analysis to include all OPEC+ members and major oil producers, integrate macroeconomic fundamentals and geopolitical risk indicators as explanatory variables, and explore ensemble forecasting methods to combine predictions from multiple network specifications. These extensions would provide a more comprehensive understanding of systemic risk transmission in global energy markets and advance the application of network-based volatility models to broader financial settings.
  
\theendnotes

 \appendix

  \clearpage

\section{Univariate ARCH and GARCH-type Model}
\label{subsec41} 

This Appendix defines the univariate ARCH and GARCH models employed as building blocks in the CCC, DCC, and GO-GARCH frameworks. We also introduce the univariate log-ARCH process, which later serves as the foundation for the network-based log-volatility models.

\subsection{Univariate ARCH and GARCH process}
\label{subsec411} 

Since the seminal work of \citet{engleAutoregressiveConditionalHeteroscedasticity1982}, autoregressive conditional heteroscedastic (ARCH) models have become central to econometric forecasting, allowing time-varying volatility to be explicitly modeled in financial time series. \citet{bollerslevGeneralizedAutoregressiveConditional1986} extended this framework through the Generalized ARCH (GARCH) model, where volatility depends on past squared observations and previous conditional variances. Both the ARCH and GARCH models capture conditional variance dynamics separately from the conditional mean, which may follow temporal trends, seasonal patterns, or autoregressive structures. In contrast, stochastic volatility models introduce a latent process to govern volatility evolution, which is also influenced by lagged squared returns. In what follows, we briefly review key univariate and multivariate versions of both classes. Let \( \{Y_t \in \mathbb{R}^r : t \in \mathbb{Z} \} \) denote a discrete-time, \( r \)-dimensional random process observed at equally spaced intervals.

For a univariate time series (i.e., \( r = 1 \)), the process \( Y_t \) is given by
\begin{equation}
	Y_t = h_t^{1/2} \varepsilon_t\,,
\end{equation}
where $h_{t}^{1/2}$ scales the i.i.d. sequence \( \{ \varepsilon_t \} \) with zero mean and unit variance, and \( h_t \) denotes the conditional variance of \( Y_t \), interpreted as time-varying volatility. In volatility models, \( h_t \) evolves based on past realizations of the process.

In an ARCH(\( p \)) model, the volatility dynamics are specified as
\begin{equation}
	h_t = \alpha_0 + \sum_{i=1}^p \alpha_i Y_{t-i}^2,
\end{equation}
while the GARCH(\( p, q \)) specification is given by
\begin{equation} \label{eq:univ_garch}
	h_t = \alpha_0 + \sum_{i=1}^p \alpha_i Y_{t-i}^2 + \sum_{j=1}^q \beta_j h_{t-j},
\end{equation}
where \( \alpha_0 > 0 \), \( \alpha_i \geq 0 \), and \( \beta_j \geq 0 \) ensure the positivity of the conditional variance. The process is weakly stationary if \( \sum_{i=1}^p \alpha_i + \sum_{j=1}^q \beta_j < 1 \). For a comprehensive treatment of GARCH-type models, see \citet{francq2019garch}.

\subsection{Log-ARCH and log-GARCH models}
\label{subsec42}

The log-volatility model was first proposed by \citet{geweke1986commet} and later supported by \citet{pantula1986comment} and \citet{milhoj1987multiplicative}. It avoids non-negativity constraints on the coefficients and facilitates the inclusion of regressors in the volatility specification. A log-ARCH model is given by
\begin{equation}
	\log h_{t} = \alpha_0 + \sum_{i=1}^{p} \alpha_i \log Y_{t-i}^2.
\end{equation}

This can be extended to a log-GARCH model:
\begin{equation}
	\log h_{t} = \alpha_0 + \sum_{i=1}^{p} \alpha_i \log Y_{t-i}^2 + \sum_{j=1}^{q} \beta_j \log h_{t-j}.
\end{equation}

Log-GARCH models exhibit multiplicative volatility dynamics, unlike the additive structure of standard ARCH and GARCH models \citep[see][]{francq2013garch}. In these models, volatility is a deterministic function of past observations. In contrast, stochastic volatility (SV) models treat volatility as a latent process. A standard SV model specifies log-volatility as an AR(1) process:
\begin{equation}
	\log h_t - \mu_h = \phi(\log h_{t-1} - \mu_h) + u_t,
\end{equation}
where \( |\phi| < 1 \), and \( \varepsilon_t \) and \( u_t \) are independent shocks. For details on the statistical properties of the SV models, see \citet{ghysels1996statistical} and \citet{shephard2005stochastic}.

 \section{Supplementary Distributional Analysis}
 \label{app:distribution}
 
 \renewcommand{\thefigure}{B.\arabic{figure}}
 \setcounter{figure}{0}
 
 This appendix provides a detailed visual examination of the distributional characteristics of the monthly oil price log returns for the selected OPEC members.
 
\begin{figure}[!ht] 
 	\centering
 	\includegraphics[width=0.77\textwidth]{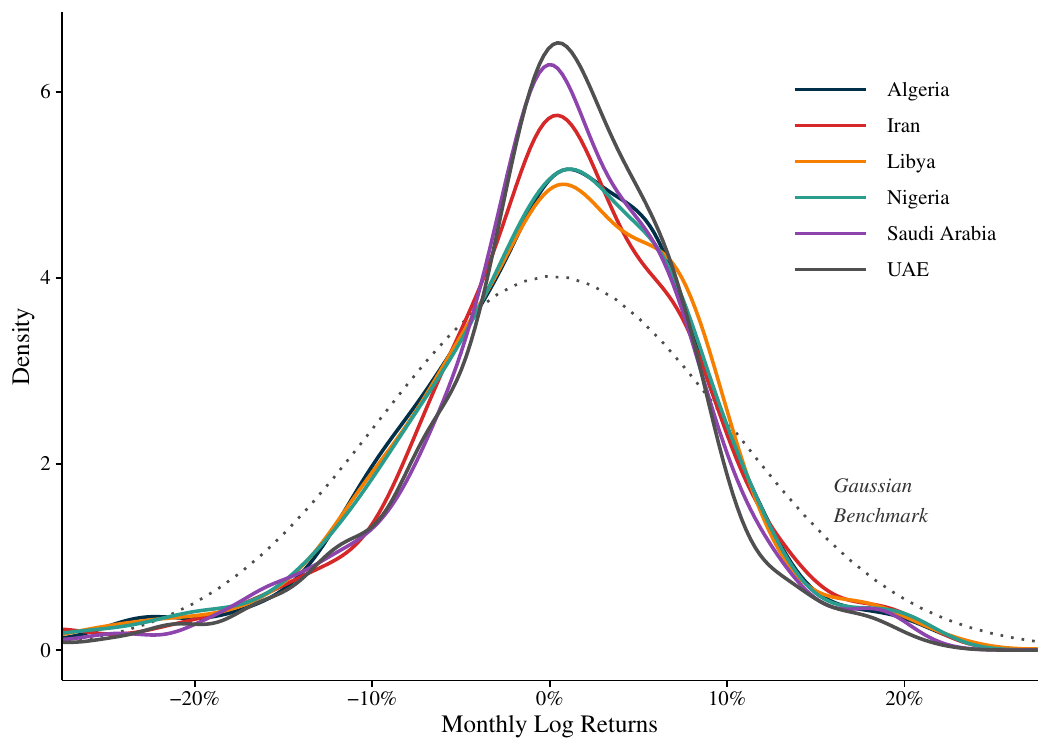} 
 	\caption{Kernel Density Estimates vs. Gaussian Benchmark}
 	\label{fig_app:density_plots}
 	\vspace{0.1cm} 
 	\begin{minipage}{\textwidth}
 		\footnotesize
 		\textit{Notes:} The figure displays the kernel density estimates of monthly log returns for each country (solid colored lines). The dotted black curve represents a theoretical Normal distribution with mean and standard deviation equal to the sample moments. The sharp peaks around the mean and the extended tails in the empirical distributions highlight the leptokurtic nature of oil price returns, justifying the use of heavy-tailed distributions in the GARCH modeling framework.
 	\end{minipage}
 \end{figure}

 Figure \ref{fig_app:density_plots} illustrates the kernel density estimates for the monthly log returns of all six countries. The distributions are overlaid with a Gaussian (Normal) benchmark (dotted line) derived from the aggregate sample moments. All countries exhibit significant leptokurtosis (peakedness) and fat tails relative to the normal distribution, visually confirming the non-normality documented in the descriptive statistics.
 
\begin{figure}[!ht] 
 	\centering
 	\includegraphics[width=0.76\textwidth]{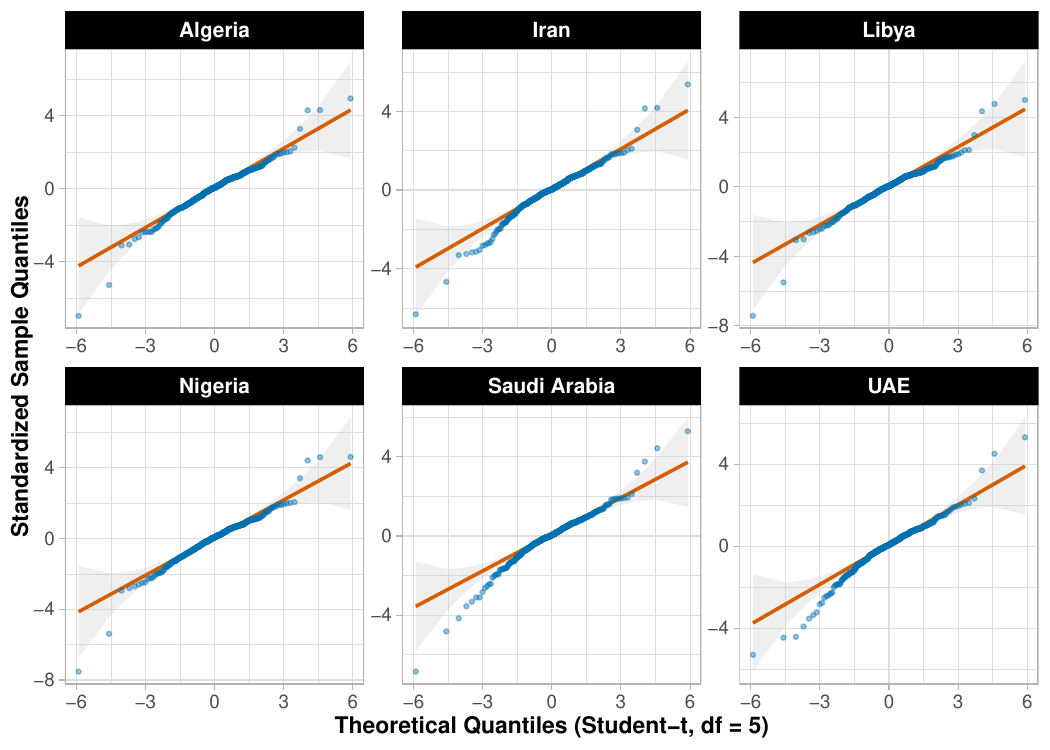} 
 	\caption{Quantile-Quantile Plots: Standardized Returns vs. Student-$t$ Distribution}
 	\label{fig_app:qq_plots}
 	\vspace{0.1cm} 
 	\begin{minipage}{\textwidth}
 		\footnotesize
 		\textit{Notes:} Each panel displays the Q-Q plot for a specific OPEC member country. The solid red line represents the theoretical Student-$t$ distribution with 5 degrees of freedom. The shaded gray region indicates the 95\% confidence interval. Points lying outside the confidence bands suggest departures from the assumed distributional form. Returns are standardized within each country (mean = 0, standard deviation = 1) to ensure comparability across panels.
 	\end{minipage}
 \end{figure}

 Figure \ref{fig_app:qq_plots} further investigates these deviations by presenting Quantile-Quantile (Q-Q) plots. We compare the empirical distribution of standardized returns against a theoretical Student-$t$ distribution with 5 degrees of freedom.
 
\renewcommand{\thetable}{C.\arabic{table}}
\setcounter{table}{0}
\section{Estimation of the Univariate GARCH models}

\begin{table}[H] 
	\centering
	\small 
	\caption{Univariate GARCH(1,1) estimates with Student’s t-innovations for monthly log-returns}
	\label{tab_app:garch_results}
	\begin{tabular}{lcccccc}
		\toprule
		\textbf{Parameter} & \textbf{Algeria} & \textbf{Iran} & \textbf{Libya} & \textbf{Nigeria} & \textbf{S. Arabia} & \textbf{UAE} \\
		\midrule
		$\mu$ & 0.0025 & 0.0058 & 0.0027 & 0.0016 & 0.0046 & 0.0045 \\
		& (0.4437) & (0.0636) & (0.4207) & (0.6514) & (0.2346) & (0.1988) \\[0.5ex]
		$\omega$ & 0.0008 & 0.0013 & 0.0007 & 0.0006 & 0.0008 & 0.0007 \\
		& (0.0762) & (0.0264) & (0.1680) & (0.2940) & (0.3893) & (0.3556) \\[0.5ex]
		$\alpha_1$ & 0.4210 & 0.5556 & 0.4335 & 0.4389 & 0.5191 & 0.4882 \\
		& (0.0000) & (0.0000) & (0.0000) & (0.0000) & (0.0000) & (0.0000) \\[0.5ex]
		$\beta_1$ & 0.5549 & 0.4161 & 0.5655 & 0.5601 & 0.4799 & 0.5108 \\
		& (0.0000) & (0.0001) & (0.0000) & (0.0000) & (0.0039) & (0.0013) \\[0.5ex]
		$\nu$ & 8.7223 & 5.6097 & 9.0256 & 9.3528 & 5.2138 & 5.3147 \\
		& (0.0019) & (0.0001) & (0.0021) & (0.0026) & (0.0001) & (0.0000) \\[0.5ex]
		LogLik & 541.35 & 546.58 & 533.48 & 540.81 & 581.23 & 609.66 \\
		\bottomrule
	\end{tabular}
	\smallskip 
	\begin{minipage}{\linewidth}
		\footnotesize \centering
		\textbf{Notes:} p-values are reported in parentheses. $\nu$ denotes the degrees of freedom of the Student’s t-distribution.
	\end{minipage}
\end{table}

\vspace{-0.4cm} 

\section{Computational Performance Comparison}
\renewcommand{\thetable}{D.\arabic{table}}
\setcounter{table}{0}

\begin{table}[!ht] 
	\centering
	\caption{Computational Performance Comparison: Standard MGARCH vs Network log-ARCH Models}
	\label{tab_app:computational_comparison}
	\resizebox{\textwidth}{!}{%
		\begin{tabular}{c|ccc|cccccc}
			\hline\hline
			& \multicolumn{3}{c|}{\textbf{Standard MGARCH Models}} & \multicolumn{6}{c}{\textbf{Network log-ARCH Models}} \\
			\cline{2-10}
			\textbf{T0} & \textbf{CCC} & \textbf{DCC} & \textbf{GO} & \textbf{Euclidean} & \textbf{Correlation} & \textbf{Piccolo} & \textbf{Net. CCC} & \textbf{Net. DCC} & \textbf{Net. GO} \\
			\hline
			\multicolumn{10}{c}{\textit{Panel A: Average Forecasting Time (seconds per step)}} \\
			\hline
			200 & 0.578 & 0.989 & 0.153 & 0.000022 & 0.000018 & 0.000017 & 0.000017 & 0.000018 & 0.000017 \\
			250 & 0.591 & 1.064 & 0.153 & 0.000023 & 0.000029 & 0.000017 & 0.000023 & 0.000019 & 0.000016 \\
			300 & 0.616 & 1.001 & 0.155 & 0.000022 & 0.000018 & 0.000017 & 0.000017 & 0.000017 & 0.000016 \\
			350 & 0.616 & 1.015 & 0.196 & 0.000022 & 0.000019 & 0.000017 & 0.000018 & 0.000017 & 0.000018 \\
			400 & 0.660 & 1.055 & 0.213 & 0.000023 & 0.000019 & 0.000017 & 0.000017 & 0.000018 & 0.000016 \\
			450 & 0.701 & 1.049 & 0.218 & 0.000024 & 0.000019 & 0.000017 & 0.000017 & 0.000018 & 0.000017 \\
			\hline
			\textbf{Average} & \textbf{0.627} & \textbf{1.029} & \textbf{0.181} & \textbf{0.000023} & \textbf{0.000020} & \textbf{0.000017} & \textbf{0.000018} & \textbf{0.000018} & \textbf{0.000017} \\
			\hline
			\multicolumn{10}{c}{\textit{Panel B: Average Peak RAM Usage (MB)}} \\
			\hline
			200 & 68,537 & 68,589 & 65,736 & 34,023 & 34,027 & 34,031 & 34,034 & 34,038 & 34,042 \\
			250 & 70,017 & 70,054 & 66,938 & 34,474 & 34,479 & 34,484 & 34,488 & 34,493 & 34,497 \\
			300 & 71,523 & 71,549 & 68,087 & 34,979 & 34,985 & 34,990 & 34,995 & 35,001 & 35,006 \\
			350 & 72,927 & 72,923 & 69,197 & 35,512 & 35,519 & 35,525 & 35,531 & 35,538 & 35,544 \\
			400 & 74,274 & 74,223 & 70,469 & 36,059 & 36,067 & 36,074 & 36,081 & 36,089 & 36,096 \\
			450 & 75,460 & 75,483 & 72,020 & 36,628 & 36,636 & 36,645 & 36,652 & 36,661 & 36,669 \\
			\hline
			\textbf{Average} & \textbf{72,123} & \textbf{72,137} & \textbf{68,741} & \textbf{35,279} & \textbf{35,286} & \textbf{35,292} & \textbf{35,297} & \textbf{35,303} & \textbf{35,309} \\
			\hline\hline
		\end{tabular}%
	}
	\smallskip
	\begin{minipage}{\linewidth}
		\scriptsize 
		\textit{Note:} All models use a tolerance parameter of $\epsilon = 1.006 \times 10^{-8}$. Panel A shows the average time in seconds required to produce a single one-step-ahead volatility forecast. Panel B displays the average peak memory usage in megabytes. Network log-ARCH models demonstrate dramatically superior computational efficiency, achieving forecast speeds approximately 27,000--62,000 times faster than standard MGARCH models while using roughly 51\% less memory. CCC: Constant Conditional Correlation; DCC: Dynamic Conditional Correlation; GO: Generalized Orthogonal; Net\_CCC/DCC/GO: Network-augmented variants. Computational results were obtained using an Apple Silicon M3 processor and the R programming environment.
	\end{minipage}
\end{table}

 \renewcommand{\thetable}{E.\arabic{table}}
 \setcounter{table}{0}
 
\begin{landscape}

	\vspace*{-1em} 
	
	\section{Forecast Sensitivity Analysis}
	\label{appendix:sensitivity_table}
	
	\begin{table}[h!]
		\caption{Forecast-Error Magnitudes and Sensitivity Checks}
		\label{tab_app:accuracy_sensitivity}
		
			\centering 
			
			\setlength{\tabcolsep}{4pt} 
			\renewcommand{\arraystretch}{1.4}
			
			\begin{tabular}{lrr rrrr rrr}
				\toprule
				\multirow{2}{*}{\large Model}
				& \multicolumn{2}{c}{\large Baseline ($T_0=450$)}
				& \multicolumn{4}{c}{\large Training Window ($T_0$)}
				& \multicolumn{3}{c}{\large Zero-Adjustment ($\varepsilon$)} \\
				\cmidrule(lr){2-3}\cmidrule(lr){4-7}\cmidrule(lr){8-10}
				& \large RMSFE  & \large MAFE
				& \large 200    & \large 250    & \large 350    & \large 400
				& \large min    & \large 1\%ile & \large 1e-6 \\
				\midrule
				\multicolumn{10}{l}{\textbf{Network log-ARCH Models (with different weight matrices)}} \\
				\large Euclidean-Diss    & 2.7424 & 2.1112 & 3.2235 & 3.0711 & 2.8461 & 2.9360 & 2.7687 & 2.7424 & 2.7437 \\
				\large Correlation-Diss  & 2.7462 & 2.1166 & 3.1284 & 3.0369 & 2.7950 & 2.8476 & 2.7725 & 2.7462 & 2.7475 \\
				\large Piccolo-Diss      & 4.4773 & 3.9880 & 3.4166 & 3.5491 & 4.2584 & 6.9031 & 4.4984 & 4.4773 & 4.4783 \\
				\large Net CCC-GARCH     & 2.3267 & 1.8512 & 2.8274 & 2.6969 & 2.5425 & 2.5365 & 2.3550 & 2.3267 & 2.3281 \\
				\large Net DCC-GARCH     & 2.3268 & 1.8512 & 2.8259 & 2.6955 & 2.5426 & 2.5365 & 2.3552 & 2.3268 & 2.3282 \\
				\large Net GO-GARCH      & \textbf{2.3257} & 1.8514
				& 2.8215 & 2.6932 & \textbf{2.5422} & 2.5349
				& \textbf{2.3541} & \textbf{2.3257} & \textbf{2.3271} \\
				\midrule
				\multicolumn{10}{l}{\textbf{Standard Multivariate GARCH Models}} \\
				\large Std. CCC-GARCH    & 2.5167 & 1.7831 & 2.7350 & 2.7583 & 2.8207 & 2.5669 & 2.5167 & 2.5167 & 2.5167 \\
				\large Std. DCC-GARCH    & 2.3796 & \textbf{1.6661}
				& \textbf{2.5726} & \textbf{2.5342} & 2.5439 & \textbf{2.4116}
				& 2.3796 & 2.3796 & 2.3796 \\
				\large Std. GO-GARCH     & 6.2708 & 5.9035 & 6.3449 & 6.3171 & 6.3152 & 6.1151 & 6.2656 & 6.2708 & 6.2626 \\
				\bottomrule
			\end{tabular}
			\vspace{1ex} 
			
			\begin{minipage}{\textwidth}
				\small
				\textbf{Notes:} Baseline uses $T_0=450$ and $\varepsilon$ set to the 1\% quantile of non-zero squared returns.
				Bold entries indicate the best-performing model (lowest error) in each column.
				While Std. DCC-GARCH minimizes MAFE, the Net GO-GARCH model achieves the lowest RMSFE in the extended baseline window ($T_0=450$).
			\end{minipage}
		\end{table}
	\end{landscape}

    \section{Network Weights under Different In-Sample sizes}
    
    \renewcommand{\thefigure}{F.\arabic{figure}}
    \setcounter{figure}{0}
    
    \begin{figure}[!ht] 
    	\centering

    	\begin{minipage}[b]{0.44\textwidth} 
    		\centering
    		\includegraphics[width=\linewidth, trim={0.5cm 0cm 0.5cm 0.5cm}, clip]{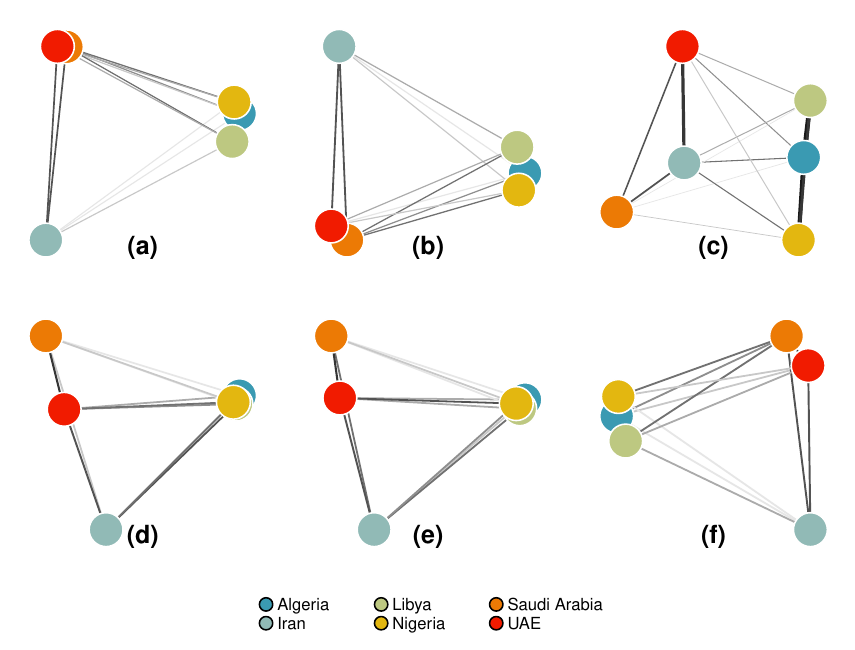}
    		\\[3pt]
    		$T_0 = 200$
    	\end{minipage}
    	\hfill
    	\begin{minipage}[b]{0.44\textwidth} 
    		\centering
    		\includegraphics[width=\linewidth, trim={0.5cm 0cm 0.5cm 0.5cm}, clip]{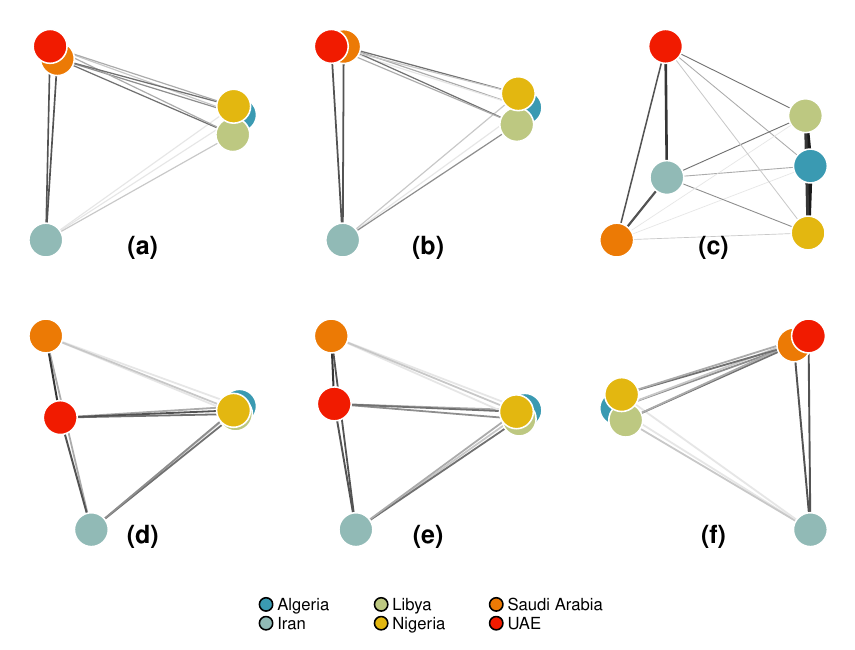}
    		\\[3pt]
    		$T_0 = 250$
    	\end{minipage}
    	
    	\vspace{0.4cm} 
    	
    	\begin{minipage}[b]{0.44\textwidth} 
    		\centering
    		\includegraphics[width=\linewidth, trim={0.5cm 0cm 0.5cm 0.5cm}, clip]{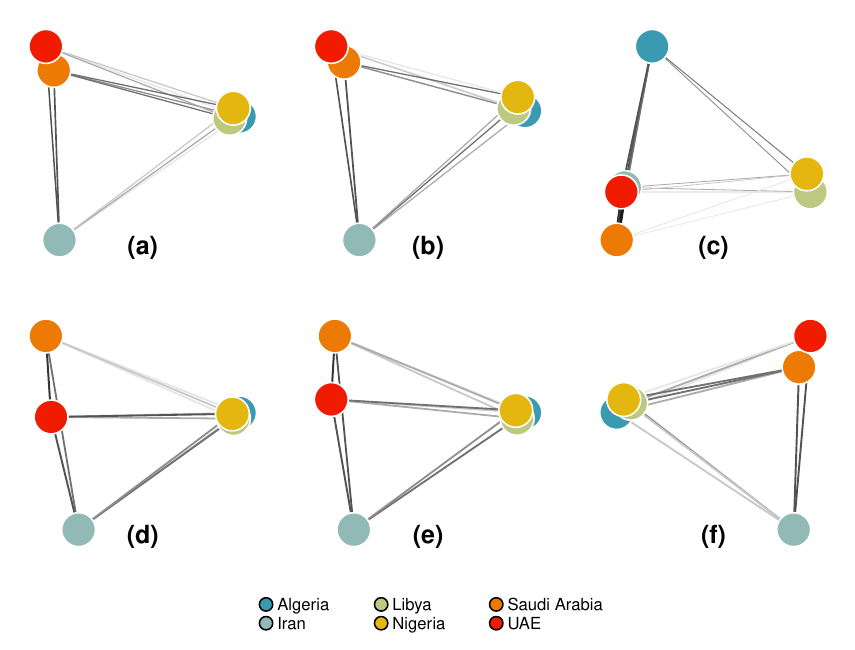}
    		\\[3pt]
    		$T_0 = 300$
    	\end{minipage}
    	\hfill
    	\begin{minipage}[b]{0.44\textwidth} 
    		\centering
    		\includegraphics[width=\linewidth, trim={0.5cm 0cm 0.5cm 0.5cm}, clip]{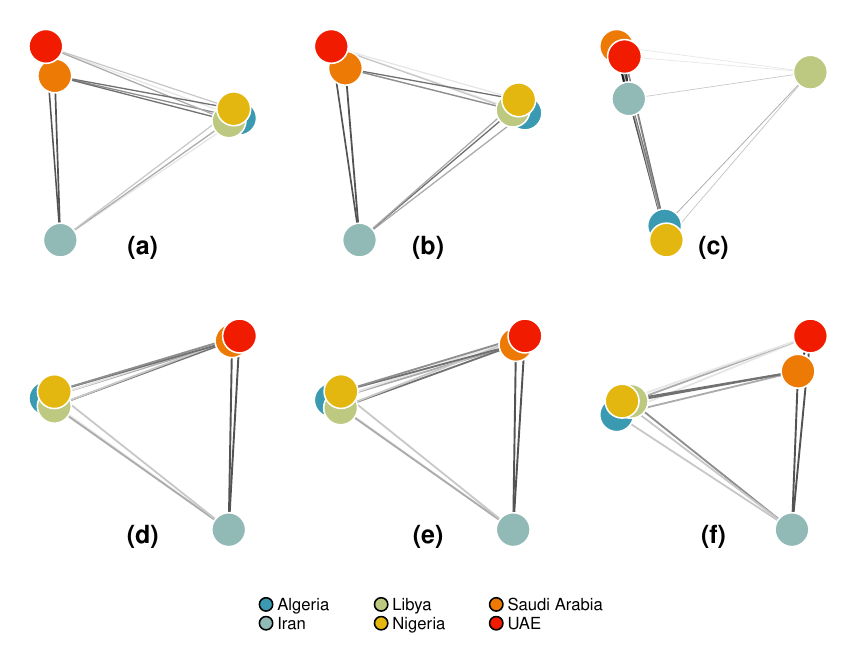}
    		\\[3pt]
    		$T_0 = 350$
    	\end{minipage}
    	
    	\vspace{0.4cm} 
    	
    	\begin{minipage}[b]{0.44\textwidth} 
    		\centering
    		\includegraphics[width=\linewidth, trim={0.5cm 0cm 0.5cm 0.5cm}, clip]{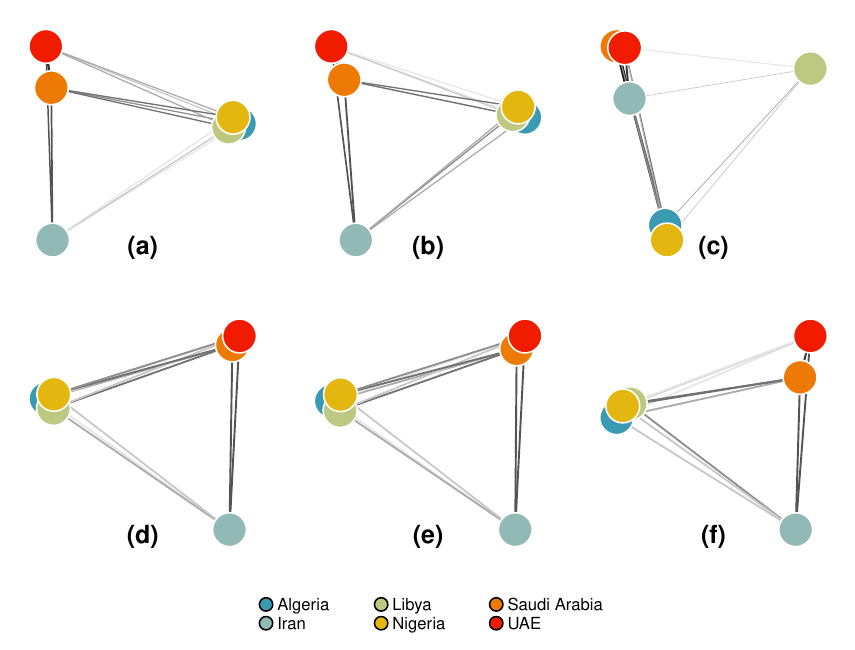}
    		\\[3pt]
    		$T_0 = 400$
    	\end{minipage}
    	\hfill
    	\begin{minipage}[b]{0.44\textwidth} 
    		\centering
    		\includegraphics[width=\linewidth, trim={0.5cm 0cm 0.5cm 0.5cm}, clip]{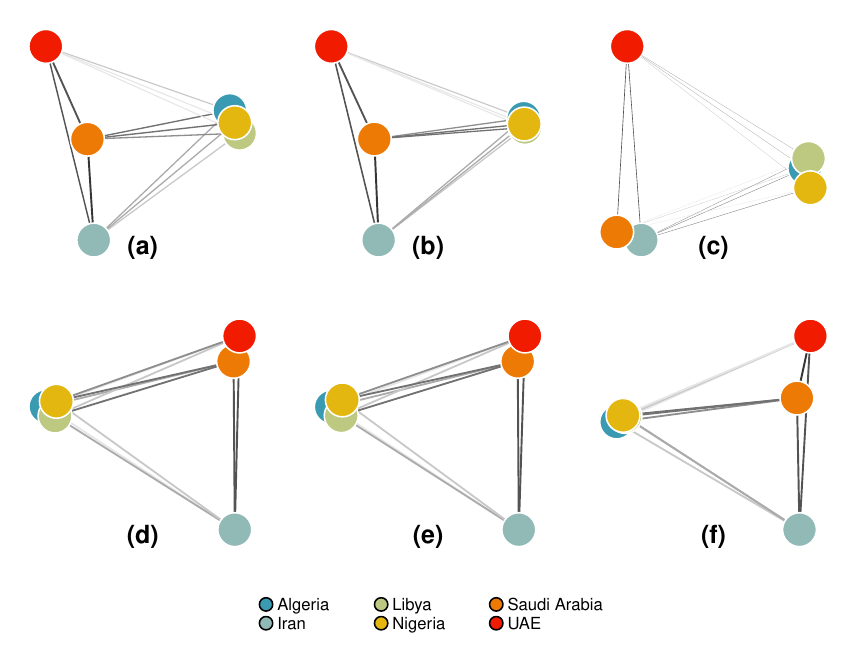}
    		\\[3pt]
    		$T_0 = 450$
    	\end{minipage}
    	
    	\caption[Evolution of Network Weights for Different In-Sample Sizes]{%
    		\textbf{Evolution of Network Weights for Different In-Sample Sizes ($T_0$).} 
    		Note: Network plots of the six OPEC oil-exporting countries based on six different distance measures. 
    		Raw data-based distances: (a) Euclidean distance, (b) Correlation-based distance. 
    		Model-based distances: (c) Piccolo-based distance, (d) CCC-GARCH, (e) DCC-GARCH, and (f) GO-GARCH. The grey colour scale of the edges is proportional to the weight degree of the connections.}
    	\label{fig_app:six_weight_matrices}
    \end{figure}

  \clearpage

\bibliographystyle{Working_paper}
\bibliography{Working_paper}

\end{document}